\documentclass[pdflatex,sn-mathphys-num]{sn-jnl}

\usepackage{graphicx}%
\usepackage{multirow}%
\usepackage{amsmath,amssymb,amsfonts}%
\usepackage{amsthm}%
\usepackage{mathrsfs}%
\usepackage[title]{appendix}%
\usepackage{xcolor}%
\usepackage{textcomp}%
\usepackage{manyfoot}%
\usepackage{booktabs}%
\usepackage{algorithm}%
\usepackage{algorithmicx}%
\usepackage{algpseudocode}%
\usepackage{listings}%

\usepackage{subfigure}
\usepackage{caption}
\usepackage{xcolor}    
\usepackage{colortbl}  

\theoremstyle{thmstyleone}%
\theoremstyle{thmstyletwo}%

\theoremstyle{thmstylethree}%

\begin{document}

\title[Uni-Mol3: A Multi-Molecular Foundation Model for Advancing Organic Reaction Modeling]{Uni-Mol3: A Multi-Molecular Foundation Model for Advancing Organic Reaction Modeling}


\author[1,2]{\fnm{Lirong} \sur{Wu}}\email{wulirong98@outlook.com}

\author[2,3]{\fnm{Junjie} \sur{Wang}}\email{1800011822@pku.edu.cn}

\author[2]{\fnm{Zhifeng} \sur{Gao}}\email{gaozf@dp.tech}

\author[2]{\fnm{Xiaohong} \sur{Ji}}\email{jixh@dp.tech}

\author[1,3]{\fnm{Rong} \sur{Zhu}}\email{rongzhu@pku.edu.cn}

\author[1]{\fnm{Xinyu} \sur{Li}}\email{lixy@aisi.ac.cn}

\author[2]{\fnm{Linfeng} \sur{Zhang}}\email{zhanglf@dp.tech}

\author*[2]{\fnm{Guolin} \sur{Ke}}\email{kegl@dp.tech}

\author*[1,4,5]{\fnm{Weinan} \sur{E}}\email{weinan@math.pku.edu.cn}

\affil*[1]{\small \orgdiv{AI for Science Institute}, \orgaddress{\city{Beijing}, \state{China}}}
\affil[2]{\small \orgdiv{DP Technology}, \orgaddress{\city{Beijing}, \state{China}}}
\affil[3]{\small \orgdiv{College of Chemistry and Molecular Engineering}, \orgname{Peking University}, \orgaddress{\city{Beijing}, \state{China}}}
\affil[4]{\small \orgdiv{School of Mathematical Sciences}, \orgname{Peking University}, \orgaddress{\city{Beijing}, \state{China}}}
\affil[5]{\small \orgdiv{Center for Machine Learning Research}, \orgname{Peking University}, \orgaddress{\city{Beijing}, \state{China}}}


\abstract{Organic reaction, the foundation of modern chemical industry, is crucial for new material development and drug discovery. However, deciphering reaction mechanisms and modeling multi-molecular relationships remain formidable challenges due to the complexity of molecular dynamics. While several state-of-the-art models like Uni-Mol2 have revolutionized single-molecular representation learning, their extension to multi-molecular systems—where chemical reactions inherently occur—has been underexplored. This paper introduces Uni-Mol3, a novel deep learning framework that employs a hierarchical pipeline for multi-molecular reaction modeling. At its core, Uni-Mol3 adopts a multi-scale molecular tokenizer (Mol-Tokenizer) that encodes 3D structures of molecules and other features into discrete tokens, creating a 3D-aware molecular language. The framework innovatively combines two pre-training stages: molecular pre-training to learn the molecular grammars and reaction pre-training to capture fundamental reaction principles, forming a progressive learning paradigm from single- to multi-molecular systems. With prompt-aware downstream fine-tuning, Uni-Mol3 demonstrates exceptional performance in diverse organic reaction tasks and supports multi-task prediction with strong generalizability. Experimental results across 10 datasets spanning 4 downstream tasks show that Uni-Mol3 outperforms existing methods, validating its effectiveness in modeling complex organic reactions. This work not only ushers in an alternative paradigm for multi-molecular computational modeling but also charts a course for intelligent organic reaction by bridging molecular representation with reaction mechanism understanding.}




\maketitle

\section{Introduction}\label{sec1}
In recent years, single-molecular foundation models, with Uni-Mol~\cite{zhou2023uni} and Uni-Mol2~\cite{ji2024uni} at the forefront, have made remarkable progress in the representation learning of individual molecules. These models exhibit exceptional capabilities in tasks ranging from predicting molecular properties and simulating molecular conformations to unraveling the intricate details of individual molecular structures, thereby significantly advancing our understanding and manipulation of isolated molecules. Nevertheless, despite their great successes in single-molecular systems, these models encounter huge challenges when directly tackling multi-molecular systems, such as molecular interaction prediction, molecular assembly prediction, and reaction modeling. Among these, organic chemical reaction modeling stands out as particularly challenging, since it not only involves intermolecular interactions but is also heavily influenced by environmental factors, e.g., reaction conditions, which cannot be handled by single-molecule models.

Organic reaction~\cite{chu2021desulfonylation,ali2024machine,oliveira2022machine,dong2022deep}, as the cornerstone of the modern chemical industry, plays an irreplaceable role in new material development, drug discovery, energy conversion, etc. At its core, organic reaction involves the directed construction of target molecules through precise regulation of chemical bond breaking and formation. Reaction selectivity control, synthetic pathway design, and reaction mechanism analysis remain the key challenges in this field. Historically dependent on expert experience and reaction templates~\cite{chen2024reaction,seidl2022improving,wang2025reacon}, organic reaction is now undergoing a paradigm shift toward data-driven, template-free approaches~\cite{das2024advances,li2024research,zhong2024recent}, accelerated by the exponential growth of chemical data and advancements in artificial intelligence (AI) technologies. Notably, emerging AI models~\cite{schwaller2019molecular,dai2019retrosynthesis,lu2022unified,coley2017prediction,shi2020graph} have demonstrated the capacity to match or exceed human expertise in tasks such as retrosynthetic prediction, yield estimation, etc. These data-driven approaches offer innovative solutions to long-standing empirical challenges in organic reaction, thereby opening new frontiers for efficient and intelligent synthetic design.

Recent data-driven models for organic reactions can be broadly classified into three categories: descriptor-based, graph-based, and sequence-based methods. Descriptor-based methods~\cite{ahneman2018predicting,probst2022reaction,schneider2015development,karthikeyan2014representation,segler2017neural,coley2017prediction} typically leverage hand-crafted features derived from reaction templates and feature engineering as molecular representations, coupled with traditional machine learning models for downstream tasks. While effective in small-scale datasets, these models often exhibit limited generalizability and struggle with scalability to large-scale data. 
By contrast, graph-based methods~\cite{mcdermott2021graph,he2008graph,jin2017predicting,coley2019graph,dai2019retrosynthesis,shi2020graph} treat atoms as nodes and valence bonds as edges, formulating chemical reactions as atomic rearrangement processes involving bond breaking and formation. Although these models can capture reaction mechanisms at the atomic level, they rely heavily on atom-mapping information and lack a unified framework for integrating diverse reaction tasks. 
On the other hand, sequence-based methods~\cite{schwaller2019molecular,irwin2022chemformer,lu2022unified,jiang2021smiles,schwaller2021prediction,schwaller2018found,schwaller2018found,karpov2019transformer} encode molecules as text sequences, framing organic reaction tasks as language translation problems. These sequence-based models enable seamless integration of molecular representations with natural language processing techniques, empowering flexible handling of complex reaction scenarios and emerging as a mainstay in the field.

\begin{figure}[t]
\centering
    \subfigure[Top-1 Accuracy]{\includegraphics[width=0.48\linewidth]{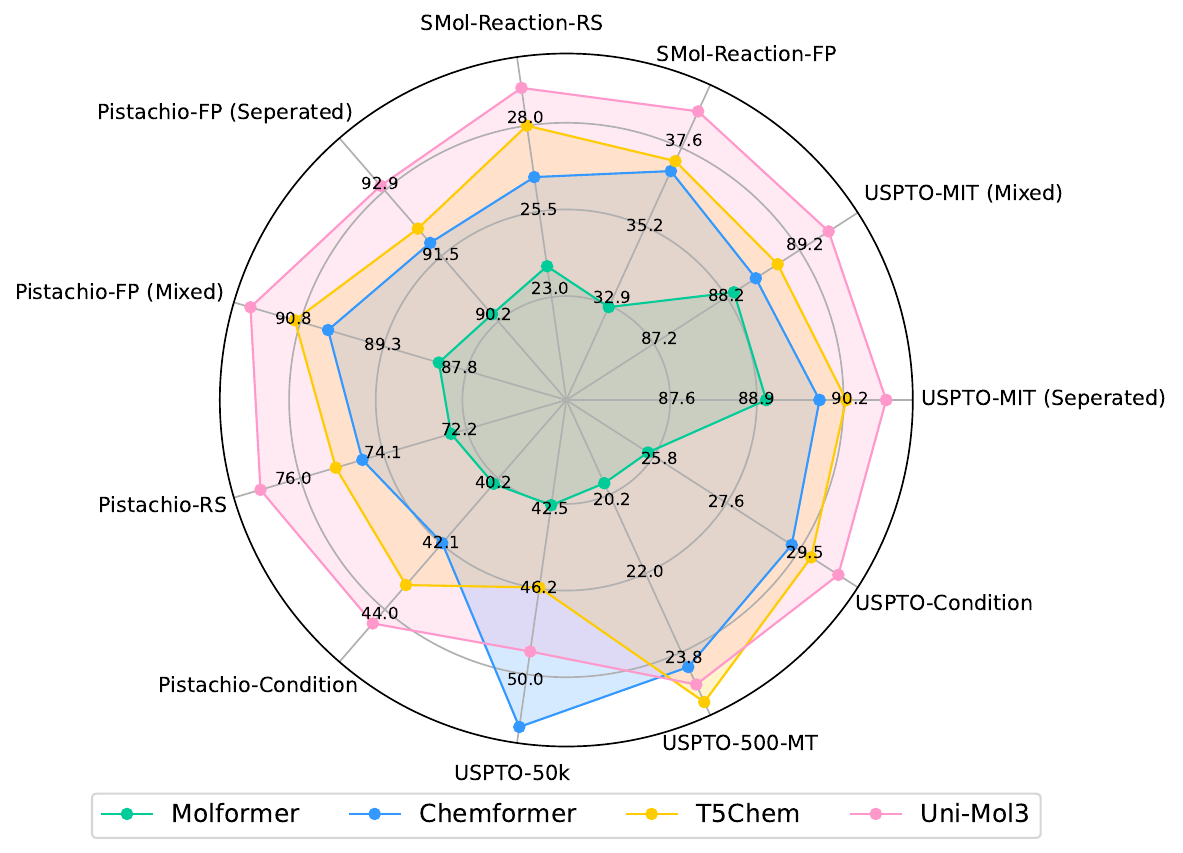}}
    \subfigure[Levenshtein Distance (LEV)]{\includegraphics[width=0.48\linewidth]{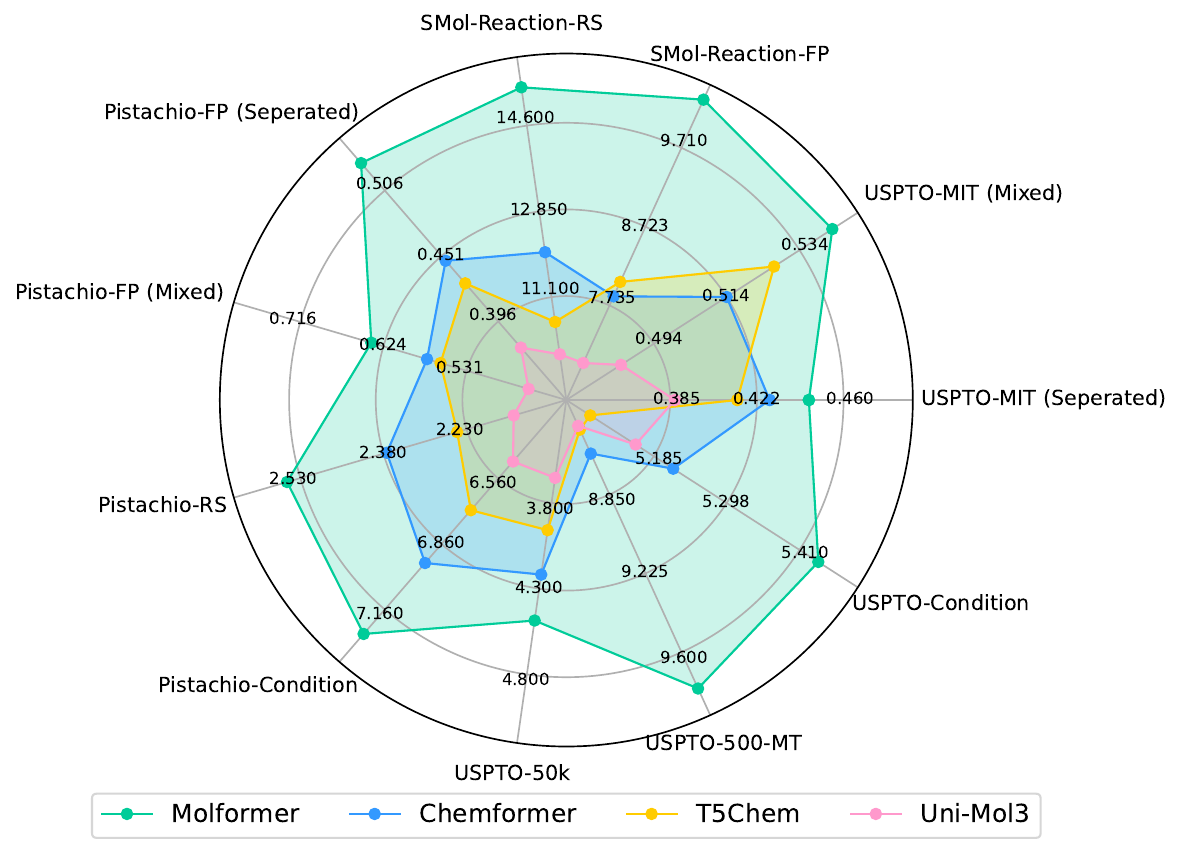}}
    \subfigure[MFP-TC]{\includegraphics[width=0.48\linewidth]{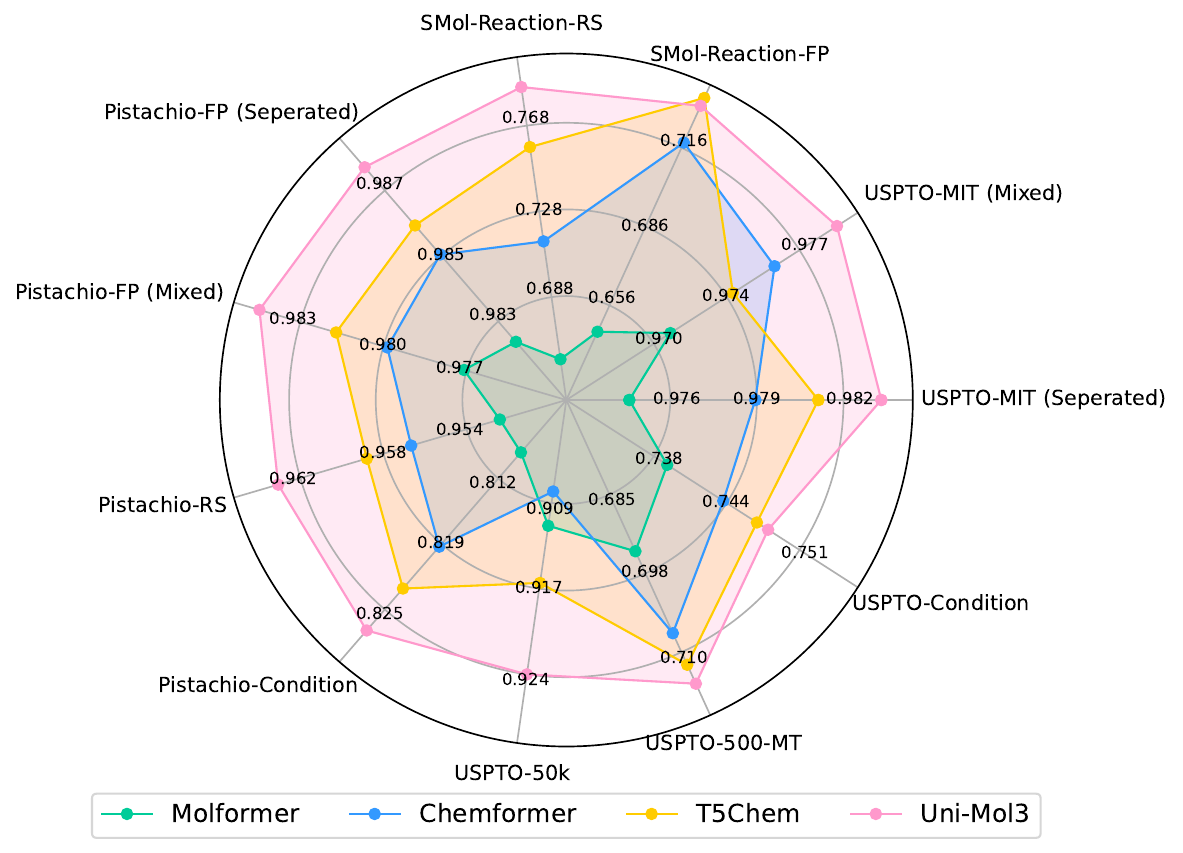}}
    \subfigure[Invalidity Rate (\%)]{\includegraphics[width=0.48\linewidth]{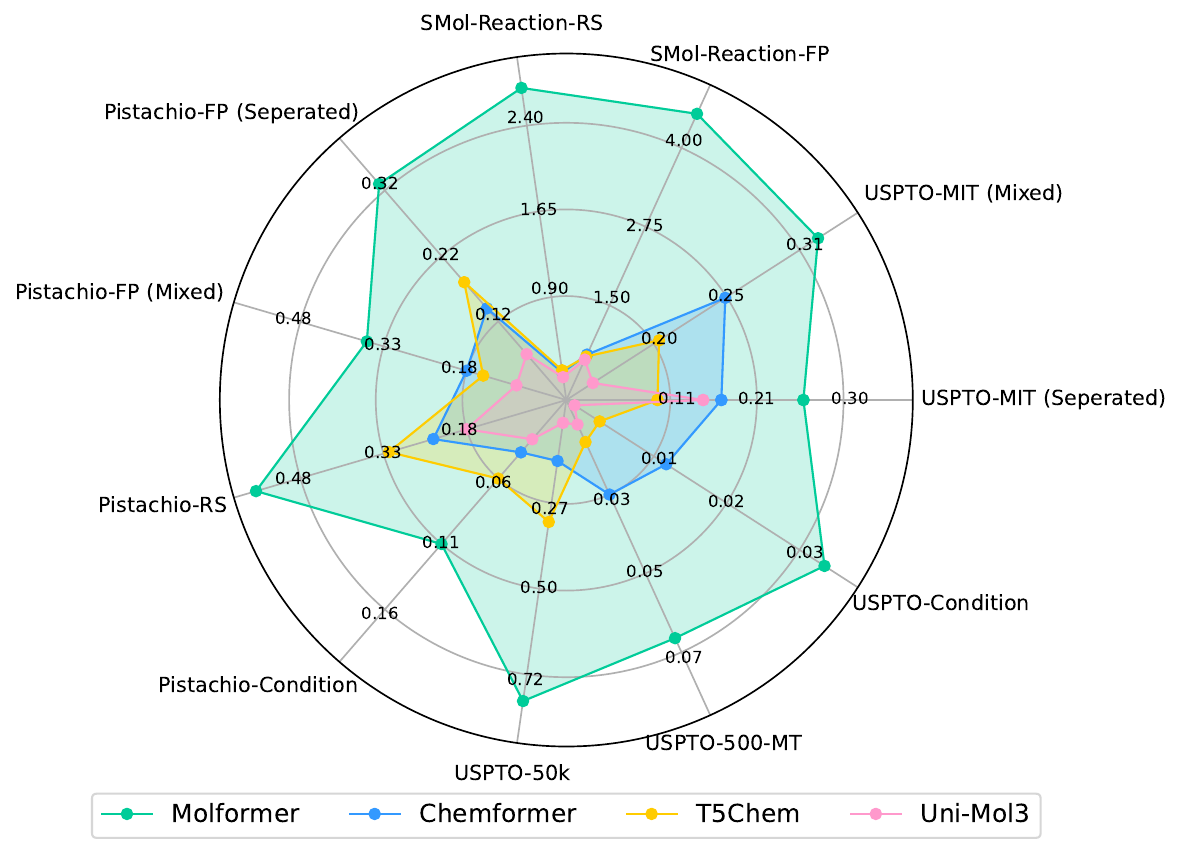}}
    \caption{Four radar plots (each corresponding to one evaluation metric) of 11 sets of experiments on multiple datasets, where Uni-Mol3 is compared to three previous baselines, including Molformer~\cite{schwaller2019molecular}, Chemformer~\cite{irwin2022chemformer}, and T5Chem~\cite{lu2022unified}. Among the four evaluation metrics, higher Top-1 accuracy and Tanimoto coefficient of molecular molar fingerprint (MFP-TC) are preferred, while lower Levenshtein distance (LEV) and invalidity rate are better. For condition generation, three datasets are used, including USPTO-500-MT, USPTO-Condition, and Pistachio-CG. Retrosynthetic prediction involves three datasets, USPTO-50k, SMol-Reactions-RS, and Pistachio-RS. As for product prediction, three datasets—USPTO-MIT, SMol-Reactions-FP, and Pistachio-FP—have two different settings for the reactant-condition \textit{“separated”} and \textit{“mixed”}, respectively. The experimental results demonstrate the overall advantages of Uni-Mol3 over other baselines across a wide range of tasks, datasets, and evaluation metrics.}
    \label{fig:radar}
\end{figure}

Despite great progress in applying Transformer architectures with SMILES~\cite{weininger1988smiles} inputs to organic reaction~\cite{schwaller2019molecular,irwin2022chemformer,lu2022unified,karpov2019transformer}, several fundamental challenges remain unresolved. Firstly, as 2D molecular descriptors, SMILES strings fail to encode full-atom 3D coordinates and stereochemical details~\cite{mislow2012introduction,nogradi2016stereochemistry,andersen2017chemical}, hindering models from capturing critical reaction features like steric hindrance and chiral induction. This compromises the intrinsic relationship between molecular spatial structure and chemical reactivity, thereby impeding accurate modeling of reaction selectivity and pathways. Secondly, organic reactions involve both single- and multi-molecular systems, where the former requires modeling intra-molecular grammars, and the latter emphasizes inter-molecular dependencies. The integration of single-molecular grammars with multi-molecular dependencies to build a unified pre-training framework remains a significant hurdle for intelligent reaction modeling. Lastly, although unified frameworks have been developed for diverse reaction tasks~\cite{lu2022unified,cao2024presto}, the scarcity of high-quality data and inconsistent annotation severely limit model generalization in complex reaction scenarios. More crucially, the absence of standardized benchmarks for cross-task evaluations and multi-dataset analyses hampers fair comparisons among models. 

Building on Uni-Mol2’s superior single-molecular representation capabilities~\cite{zhou2023uni,ji2024uni}, this work introduces Uni-Mol3, a novel deep learning framework that enables unified multi-molecular reaction modeling via a hierarchical pipeline. First, we propose a 3D structure-aware molecular language system, where a multi-scale Mol-Tokenizer quantizes 1D atomic features, 2D graph structures, and 3D coordinates into discrete tokens—addressing SMILES’ inherent limitation in capturing spatial information. Furthermore, Uni-Mol3 employs a two-tier pre-training strategy: molecular pre-training learns single-molecular grammatical rules, while subsequent reaction pre-training captures thermodynamic and kinetic principles of multi-molecular reactions, forming a progressive learning framework from molecular grammars to reaction mechanisms. With prompt-aware downstream fine-tuning, Uni-Mol3 can adapt to diverse chemical reaction tasks without or with minimal output-layer modifications. As demonstrated by the four radar charts in Figure.~\ref{fig:radar}, Uni-Mol3 outperforms state-of-the-art baselines in four evaluation metrics across multiple datasets for diverse tasks, establishing a versatile and efficient paradigm for intelligent multi-molecular reaction modeling.

\section{Methodology}

\begin{figure}[t]
\centering
    \includegraphics[width=1.0\textwidth]{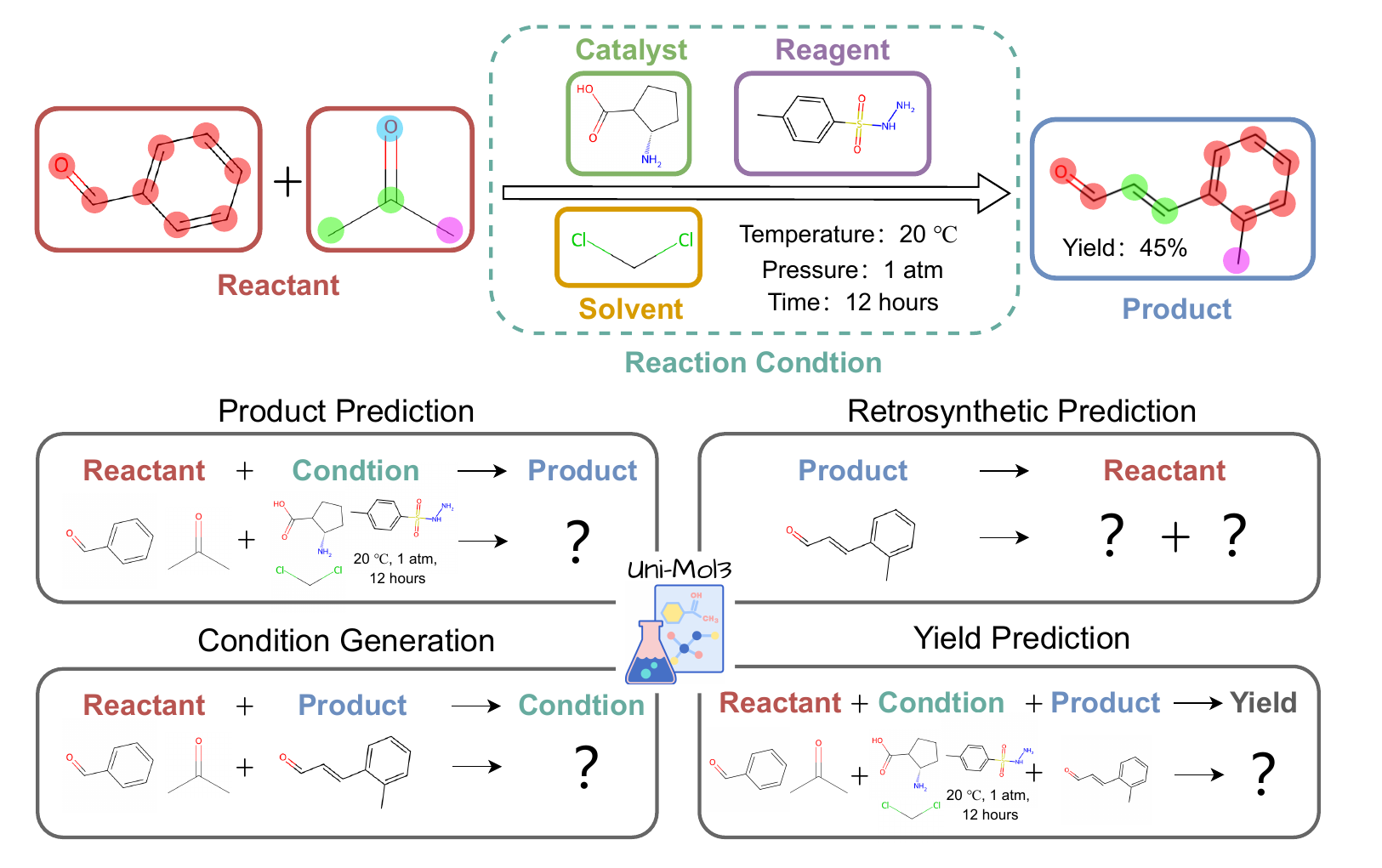}
    \caption{\textbf{Top:} Illustration of a classical chemical reaction, which involves three
elements: reactants, reaction conditions, and products. The reaction conditions consist of catalyst, solvent, reagents, temperature, pressure, reaction time, etc. \textbf{Bottom:} A high-level overview of the four representative reaction tasks supported by Uni-Mol3, which visualizes the mappings of reactants, conditions, products, and reaction yields.}
    \label{fig:reaction}
\end{figure}

\textbf{Problem Statement.} Chemical reaction is a process in which molecules are transformed by recombination of atoms, as illustrated in Figure.~\ref{fig:reaction}, and its basic components cover three elements: reactants, reaction conditions, and products. Reactants serve as the initial molecular entities and participate in the transformation through the breaking and formation of chemical bonds. Products, as the final molecular outcomes, derive their composition and structure from the atomic types of the reactants and the reaction conditions. Reaction conditions are the key variables that determine the reaction path and product yields, including temperature, pressure, catalysts, solvents, reagents, etc. Downstream tasks associated with chemical reactions include product prediction, retrosynthetic prediction, condition generation, yield prediction, etc. A wide variety of task-specific models have been developed for different tasks in recent years, each achieving remarkable success in its respective domain. However, these seemingly diverse tasks fundamentally involve learning the mapping relationships among reaction components, as shown in Figure.~\ref{fig:reaction}. This shared nature implies the potential to unify these tasks under a cohesive framework of conditional generation or regression.

Building upon the seminal advances in molecular representation learning by the prior work Uni-Mol2, we present Uni-Mol3—a novel deep learning framework that extends its foundational architectures to multi-molecular reaction modeling. Uni-Mol3 demonstrates unique versatility in addressing a broad range of chemical reaction tasks with minimal modifications, highlighting its effectiveness in unified reaction modeling.

\begin{figure}[t]
\centering
    \includegraphics[width=1.0\textwidth]{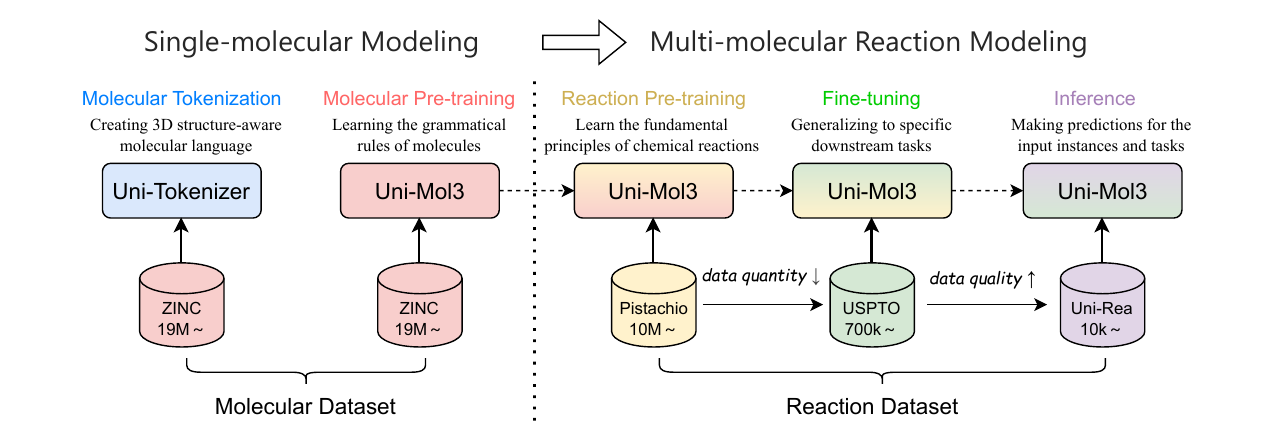}
    \caption{A high-level overview of hierarchical pipeline for Uni-Mol3, in which molecular modeling is extended from the single- to the multi-molecular systems. For single-molecular modeling, we perform molecular tokenization and pre-training to create 3D structure-aware molecular language and learn molecular grammatical rules, respectively. For multi-molecular modeling, we perform large-scale reaction pre-training to learn fundamental principles of chemical reactions, then fine-tune the model to generalize to downstream tasks, and finally make predictions for input instances and tasks.}
    \label{fig:pipeline}
\end{figure}

\begin{figure}[h]
\centering
    \includegraphics[width=1.0\textwidth]{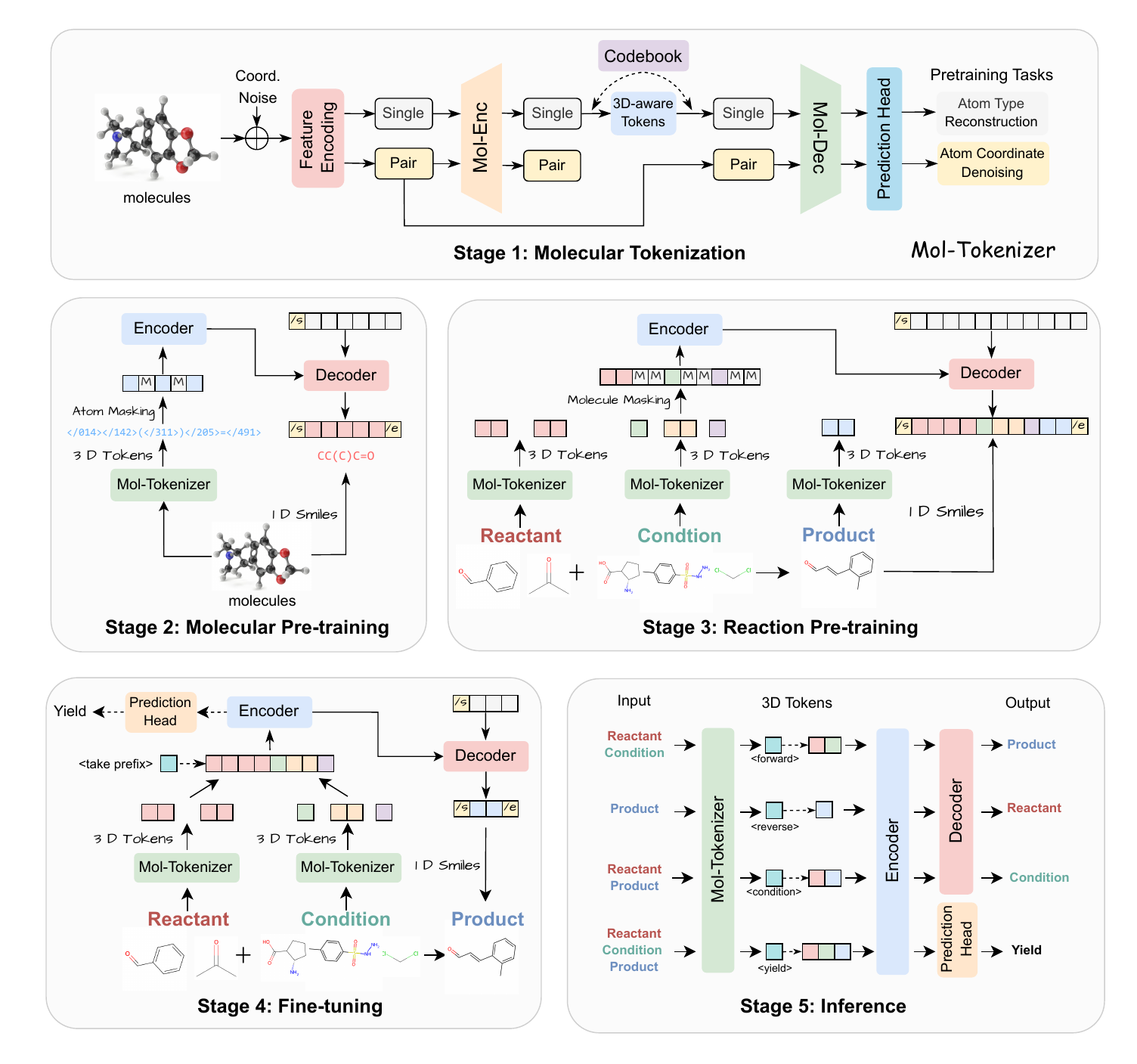}
    \caption{An overview of the five primary stages in Uni-Mol3, which extends molecular modeling from the single- to multi-molecular level, applicable to various chemical tasks.}
    \label{fig:framework}
\end{figure}

\textbf{Hierarchical Framework for Uni-Mol3.} 
Uni-Mol3 employs a hierarchical training pipeline and data organization framework, as illustrated in Figure.~\ref{fig:pipeline}. The process starts with single-molecular modeling on a large-scale molecular dataset. Leveraging the pre-trained Uni-Mol2, we train a Mol-Tokenizer to quantize multi-scale molecular information into discrete tokens, establishing a 3D structure-aware molecular language. Following previous successful practices, the encoder-decoder architecture serves as the backbone of Uni-Mol3. During molecular pre-training, atom-level masked modeling and next token prediction are used to learn single-molecular grammars. The pre-trained Uni-Mol3 is extended to multi-molecular reaction modeling via reaction pre-training with molecular-level masked modeling on large-scale reaction datasets, capturing the fundamental principles of chemical reactions. Subsequently, Uni-Mol3 is fine-tuned with task-specific prompts to enhance its generalizability for downstream reaction tasks, enabling flexible predictions for interested tasks with given inputs.

\subsection{Single-molecular Modeling}
\subsubsection{Molecular Tokenization (Stage 1)}
SMILES (Simplified Molecular Input Line Entry System)~\cite{weininger1988smiles} constitutes a compact molecular notation system that encodes essential molecular information—including atom types, bonding connectivity, and bond orders—into text strings via specific rules. Widely adopted in chemical and pharmaceutical research, its sequential nature enables straightforward textual modeling of chemical reactions, allowing language models to process chemical data efficiently without complex feature engineering~\cite{schwaller2019molecular,lu2022unified,karpov2019transformer}. However, SMILES-based language models face inherent limitations in tackling complex chemical problems, primarily due to the absence of explicit 3D structural information. The spatial arrangement of atoms and their interactions within a 3D space directly govern the feasibility, reaction rate, and selectivity of chemical transformations. For instance, in asymmetric catalytic reactions, the spatial compatibility between catalyst and substrate is critical for achieving high stereoselectivity; in enzymatic reactions, the 3D microenvironment of active sites determines substrate specificity and catalytic efficiency. These processes rely heavily on molecular 3D structures, which cannot be accurately captured by SMILES' 2D connectivity information. Thus, developing a 3D structure-aware molecular language is imperative. Such a language should integrate explicit 3D structural descriptors to capture reaction-driven molecular dynamics, while inheriting SMILES' sequential property for bridging to existing language models.

Leveraging the powerful representational capacity of Uni-Mol2 as a foundation, we propose a multi-scale molecular tokenizer (Mol-Tokenizer), as shown in Figure.~\ref{fig:framework}. Mol-Tokenizer quantizes multi-scale molecular information—1D atom types, 2D molecular graphs, and 3D conformations—into 3D structure-aware discrete tokens (abbr. 3D tokens). Next, we introduce how to construct and train the Mol-Tokenizer, including feature engineering, the encoder and decoder, quantization, and training tasks.
\newline

\textbf{Feature Encoding.} Given a molecule $M=(x,e,r)$, where $x\in\mathbb{R}^{N\times d_x}$ denotes 1D atom features, $e\in\mathbb{R}^{N\times N\times d_e}$ denotes 2D bond features, and $r\in\mathbb{R}^{N\times 3}$ denotes 3D atom coordinates. We employ RDKit to obtain atom token $x_{\text{token}}^i$, atom degree $x_{\text{degree}}^i$,
and atom types $x_{\text{type}}^i$. The single representation $x_{\text{single}}^i$ of atom $i$ is initialized as:
\begin{equation}
x_{\text {single }}^i=\operatorname{Embedding}\left(x_{\text {token }}^i\right)+\operatorname{Embedding}\left(x_{\text {degree }}^i\right)+\operatorname{Embedding}\left(x_{\text {type}}^i\right).
\end{equation}

The pair representation $x_{\text {pair}}^{i, j}$ between atom $i$ and atom $j$ is initialized as:
\begin{equation}
x_{\text {pair}}^{i, j}=\operatorname{Embedding}\left(e^{i, j}\right)+\operatorname{Embedding}\left(x_{\text {SPD }}^{i, j}\right)+x_{\text {dis }}^{i, j},
\end{equation}
where $e^{i, j}$ is the bond type, $x_{\text {SPD }}^{i, j}$ is the shortest path distance of atom pair $(i,j)$, $x_{\text {dis }}^{i, j}$ is the Euclidean distance encoded by the Gaussian kernel approach with pair type.
\newline

\textbf{Encoder and Decoder.} The encoder and decoder in Mol-Tokenizer adopt the same backbone as Uni-Mol2, each containing several two-track transformer layers. Each layer iteratively updates single and pair representations. For the $l$-th layer $\psi^{(l)}(\cdot)$, 

\begin{equation}
    h_{\text{single}}^{(l)}, h_{\text{pair}}^{(l)} = \psi^{(l)}(h_{\text{single}}^{(l-1)},h_{\text{pair}}^{(l-1)}).
\end{equation}
We initialize atom and pair embeddings $h_{\text{single}}^{(0)},h_{\text{pair}}^{(0)}$ of the first layer as $x_{\text{single}},x_{\text{pair}}$.
\newline

\textbf{Quantization.} We use FSQ (Finite Scalar Quantization)~\cite{mentzer2024finite} to quantize the continuous single representations $h_{\text{single}}^i\in\mathbb{R}^{h}$ of each atom $i$ from the encoder into a finite set of codewords $z_i$. To this end, we apply a bounding function $
f(h_{\text{single}}^{i}) = \lfloor L / 2\rfloor \tanh (h_{\text{single}}^{i})
$, and then round each channel in 
$f(x_{\text{single}}^{i})$ to a integer, as follows
\begin{equation}
    s_i = \text{round}\left(f(h_{\text{single}}^{i})\right)\in\mathbb{R}^{d},
\end{equation}
where each channel in $s_i$ takes one of $L$ unique values. Thereby, we have a codebook $s_i\in\mathcal{A}$ ($|\mathcal{A}|=L^d$) that is the product of $d$ per-channel codebook sets. The vectors in $\mathcal{A}$ can be enumerated by a simple bijection from any $s_i$ to an integer $z$ in $\{1, \cdots, L^d\}$. 
\newline

\textbf{Training.} To make discrete tokens into 3D structure-aware ones, we train the Mol-Tokenizer with two complementary tasks, i.e., atom type reconstruction and atom coordinate denoising. For the decoder, the discrete 3D token $z$ is used to initialize the single representation $\operatorname{Embedding}(z)$. To prevent leakage of atom type information, we initialize the pair representations as $x_{\text {pair}}^{i, j}$ rather than using the encoder's output $h_{\text {pair}}^{i, j}$. For atom type reconstruction, a prediction head $g_{\text{type}}(\cdot)$ directly predicts atom types from the decoder’s SINGLE representations, followed by loss computation:
\begin{equation}
\mathcal{L}_{\text{type}}=\mathcal{H}\left(x_{\text{type}},\widehat{x}_{\text{type}}\right),
\end{equation}
where $\mathcal{H}$ denotes the cross entropy, and $x_{\text{type}}$, $\widehat{x}_{\text{type}}$ are the ground-truth and predicted atom types. For the atom coordinate denoising task, we add
Gaussian noise with a standard deviation of 0.2 to all the atom coordinates $r_{\text{coor}}$. Following Uni-Mol2, we apply a position prediction head to predict the atom coordinate $\widehat{r}_{\text{coor}}$ of molecules. The losses of coordinate prediction and pairwise distance prediction are defined as
\begin{equation}
\begin{aligned}
\mathcal{L}_{\text {coor }} & =\left\|\widehat{r}_{\text {coor }}-r_{\text {coor }}\right\|_1, \\
\mathcal{L}_{\text {distance }} & =\left\|\widehat{r}_{\text {distance}}-r_{\text {distance}}\right\|_1.
\end{aligned}
\end{equation}
The final loss for training Uni-Tokenizer is summed up as:
\begin{equation}
\mathcal{L}_{\text {total }}=\mathcal{L}_{\text {type }}+\mathcal{L}_{\text {coor }}+\mathcal{L}_{\text {distance }}.
\end{equation}

\subsubsection{Molecular Pre-training  (Stage 2)}
Uni-Mol3 formulates reaction modeling as a conditional generative task, leveraging an encoder-decoder architecture to autoregressively generate target molecules. Therefore, before delving into multi-molecular reaction modeling, we first pre-train Uni-Mol3 at the single-molecular level. The objective is to enable the generative model to learn the molecular grammatical rules and chemical semantic space, thereby endowing the model with the fundamental ability to generate valid molecules. This process provides a more efficient initialization for subsequent specific tasks, such as product prediction and retrosynthesis. To this end, we first transform each input molecule $M$ into corresponding 3D tokens using the trained Uni-Tokenizer, which is defined as follows
\begin{equation}
    Z=\{z_1,z_2,\ldots,z_N\} = \text{Uni-Tokenizer}(M).
\end{equation}
Next, we sample a subset of tokens $\mathcal{M}$ from $Z$ with a ratio of 15\% and mask them with  $\textrm{[M]}$ to get $\widehat{Z}$. The encoder takes the masked 3D tokens $\widehat{Z}$ as input to generate a conditional embedding $\mathbf{c}$. The decoder generates 1D smiles $X=\{x_1,x_2,\cdots,x_N\}$ autoregressively under the condition $\mathbf{c}$ with the following optimization objective:
\begin{equation}
\mathcal{L}_{\textrm{Mol-Pre}}=-\sum_{i=1}^N \log p\left(x_i \mid x_1, x_2, \ldots, x_{i-1}, \mathbf{c} \right).
\end{equation}

\subsection{Multi-molecular Reaction Modeling}
\subsubsection{Reaction Pre-training (Stage 3)}
The chemical reaction formula can be regarded as a “chemical language” that describes the transformation process from reactants to products, such as molecular structure change, bond breaking and formation, and effects of reagents/catalysts. The objective of pre-training on chemical reactions is to enable the model to learn the syntax (reaction rules) and semantics (chemical meaning) of this “chemical language”, identify common reaction patterns, and abstract fundamental rules. This encodes a priori knowledge of chemical reactions into the model parameters, allowing the model to “understand” the reaction patterns as a chemist would. To achieve this, this subsection extends the pre-training from the single-molecular level to the multi-molecular level, with a focus on establishing the dependencies between molecules involved in a chemical reaction. Given a set of molecules $\mathcal{R}=(\mathcal{M}_R,\mathcal{M}_C,\mathcal{M}_P)$ in a chemical reaction, where $\mathcal{M}_R$, $\mathcal{M}_C$, and $\mathcal{M}_P$ represent sets of molecules for reactants, conditions (e.g., catalysts, solvents, and reagents), and products, respectively, we first transform all molecules into corresponding 3D tokens using the trained Uni-Tokenizer, as follows
\begin{equation}
    Z_i = \text{Uni-Tokenizer}(M_i), \quad\quad \forall M_i \in \mathcal{R}.
\end{equation}
The 3D tokens of all molecules can be spliced into a reaction sequence $Z_{\text{Reac}}$:
\begin{equation}
    Z_{\textrm{Reac}} = \left[Z_1,Z_2,Z_3,\cdots,Z_{|\mathcal{R}|}\right].
\end{equation}
Unlike molecular pre-training, we sample molecules from $\mathcal{R}$ rather than atoms at a ratio of 15\% and mask them with $\textrm{[M]}$ to get a masked reaction sequence $\widehat{Z}_{\textrm{Reac}}$:
\begin{equation}
    \widehat{Z}_{\textrm{Reac}} = \left[\textrm{[M]},Z_2,\textrm{[M]},\cdots,Z_{|\mathcal{R}|}\right].
\end{equation}
The masked reaction sequence $\widehat{Z}_{\text{Reac}}$ is then fed to the encoder to generate a conditional embedding $\mathbf{c}$ to be passed to the decoder. The decoder autoregressively generates 1D smile strings of the chemical reaction, and the pre-training loss is defined as
\begin{equation}
\begin{small}
\begin{aligned}
\mathcal{L}_{\textrm{Reac-Pre}}=-\sum_{i=1}^{|\mathcal{R}|}\sum_{j=1}^{N_i} \log p\left(x_{i,j} \mid x_{i,1}, x_{i,2}, \ldots, x_{i,j-1}, \{x_{i-1,k}\}_{k=1}^{N_{i-1}},\ldots,\{x_{1,k}\}_{k=1}^{N_{1}}, \mathbf{c} \right),
\end{aligned}
\end{small}
\end{equation}
where $\mathbf{c}$ is the conditional embedding, $x_{i,j}$ is the $j$-th character in the 1D smile of the $i$-th molecule, and $N_i$ is the sequence length of the 1D smile of the $i$-th molecule.

\subsubsection{Fine-tuning and Inference (Stage 4\&5)}
The objective of downstream fine-tuning is to transfer the pre-trained knowledge of Uni-Mol3 to specific downstream tasks. Task-specific prediction heads are employed based on distinct downstream types. For generative tasks, such as product prediction, retrosynthetic prediction, and condition generation, the decoder will be directly used to generate the target molecules in an autoregressive manner. For regression or classification tasks, a separate prediction head is used to predict the targets from the encoder's output. To distinguish between different tasks, a task-specific prefix token is added as a prompt to the front of the 3D token sequence output by the Uni-Tokenizer. During downstream fine-tuning, masking operations or structural noise are not introduced. This paper mainly  focuses on four representative downstream reaction tasks:
\begin{itemize}
    \item \textbf{Product Prediction.} This task takes reactants and reaction conditions as inputs, using the decoder to autoregressively generate products. It comprises two subtasks. One is labeled with the prompt token $<\!\textit{forward-sep}\!>$, where reactants and conditions are input as separate entities. The other is labeled with the prompt token $<\!\textit{forward-mixed}\!>$, indicating that reactants and conditions are mixed in input.
    \vspace{0.5em}
    \item \textbf{Retrosynthetic Prediction.} Taking products as input, this task uses the decoder to generate corresponding reactants, denoted by the prompt token $<\!\textit{reverse}\!>$.
    \vspace{0.5em}
    \item \textbf{Condition Generation.} Given reactants and products as inputs, this task uses the decoder to generate reaction conditions, denoted by prompt token $<\!\textit{condition}\!>$.
    \vspace{0.5em}
    \item \textbf{Yield Prediction.} This task takes reactants, conditions, and products as inputs and predicts the yield using a regression head, denoted by prompt token $<\!\textit{yield}\!>$.
\end{itemize}

By leveraging the fine-tuned Uni-Mol3 model, we can flexibly make inferences for the interested task with given inputs. Specifically, Uni-Tokenizer is first employed to transform input molecules into 3D token sequences, which are then concatenated with task-specific prompt tokens and fed into the encoder. Next, it selects the corresponding decoder or prediction head based on the task to be solved to generate the target output.

\section{Experiments}

\subsection{Datasets and Evaluation Metrics}
Single-molecular modeling, including Mol-Tokenizer and molecular pre-training, is conducted on the Uni-Mol dataset~\cite{zhou2023uni} containing $\sim$19M molecules, primarily from the ZINC~\cite{sterling2015zinc} and Pubmed~\cite{gaulton2012chembl} databases. For reaction pre-training, we use a large-scale reaction dataset derived from the Pistachio database developed by NextMove Software. It contains 16,678,201 chemical reactions extracted from the patent literature, making it one of the most comprehensive reaction libraries with a huge diversity in terms of reaction types and complexity. For data preprocessing of the Pistachio database, we first use RDKit to validate the chemical validity of each molecule, then remove reactions where product atoms could not be mapped to reactant atoms, and filter out reactions containing molecules with more than 80 atoms. After preprocessing, we obtain a new Pistachio-full dataset containing 11,973,789 reactions. For the three different tasks of product prediction, retrosynthetic prediction, and condition generation, we split 10,000 reactions individually for testing, resulting in three distinct datasets: Pistachio-FP, Pistachio-RS, and Pistachio-CG. For the Pistachio-CG dataset, we further filter out reactions whose conditions are unavailable or unknown, resulting in fewer samples. Finally, we perform reaction pre-training of Uni-Mol3 on the Pistachio-full dataset, followed by task-specific fine-tuning on Pistachio-FP, Pistachio-RS, and Pistachio-CG, to evaluate model generalization across diverse chemical scenarios.

We leverage several small-scale publicly accessible chemical reaction datasets for fine-tuning and evaluation, with their train/test/valid splits and corresponding downstream tasks detailed in Table~\ref{tab:dataset}. For example, USPTO-MIT~\cite{jin2017predicting} and USPTO-50k~\cite{liu2017retrosynthetic}, both curated from the USPTO database, focus on product prediction and retrosynthesis prediction, respectively. SMol-Reactions-FP and SMol-Reactions-RS from PRESTO~\cite{cao2024presto} address data leakage concerns in prior works by employing a scaffold-based splitting strategy to resample test sets, constructing non-overlapping dataset partitions that challenge model generalizability. For condition generation, we use USPTO-Condition~\cite{wang2023generic} and USPTO-500-MT~\cite{lu2022unified}: the former standardizes each reaction condition to one catalyst, two reagents, and two solvents, while the latter allows flexible specification of condition number, type, and order. To evaluate reaction yield prediction, we use the Buchwald-Hartwig dataset~\cite{ahneman2018predicting}, a high-throughput experimental collection of 3955 C-N coupling reactions. Following prior work~\cite{lu2022unified}, we adopt four out-of-sample data splits where test sets include reactions with additives not present in the training data, ensuring rigorous generalization evaluation.

\begin{table*}[!tbp]
\begin{center}
\caption{The statistical information of datasets in this work.}
\vspace{0.5em}
\label{tab:dataset}
\resizebox{1.\textwidth}{!}{
\begin{tabular}{l|cccc|c}
\toprule
\textbf{Dataset} & \textbf{\# Train} & \textbf{\# Valid} & \textbf{\# Test} & \textbf{\# All} & \textbf{Downstream Task} \\ \midrule
USPTO-MIT \cite{jin2017predicting} & 407,791 & 29,915 & 39,876 & 477,582 & Product Prediction \\
SMol-Reactions-FP \cite{cao2024presto} & 116,360 & - & 943 & 117,303 & Product Prediction \\
Pistachio-FP & 11,963,789 & - & 10,000 & 11,973,789 & Product Prediction \\ \midrule
USPTO-50k \cite{liu2017retrosynthetic} & 40,022 & 5,004 & 5,004 & 50,030 & Retrosynthesis \\
SMol-Reactions-RS \cite{cao2024presto} & 128,684 & - & 1,000 & 129,684 & Retrosynthesis \\
Pistachio-RS & 11,963,789 & - & 10,000 & 11,973,789 & Retrosynthesis \\ \midrule
USPTO-500-MT \cite{lu2022unified} & 116,360 & 12,937 & 14,238 & 143,535 & Condition Generation \\
USPTO-Condition \cite{wang2023generic} & 543,854 & 67,964 & 67,992 & 679,810 & Condition Generation \\ 
Pistachio-CG & 9,668,808 & - & 7,997 & 9,676,805 & Condition Generation \\ \midrule
Buchwald-Hartwig Test1 & 3,057 & - & 898 & 3,955 & Reaction Yield Prediction \\ 
Buchwald-Hartwig Test2 & 3,055 & - & 900 & 3,955 & Reaction Yield Prediction \\ 
Buchwald-Hartwig Test3 & 3,058 & - & 897 & 3,955 & Reaction Yield Prediction \\ 
Buchwald-Hartwig Test4 & 3,055 & - & 900 & 3,955 & Reaction Yield Prediction \\ \bottomrule
\end{tabular}}

\end{center}
\end{table*} 

We use a variety of metrics to comprehensively evaluate the model's performance across different task types. For regression tasks, we consider the following metrics: (1) Mean Absolute Error (MAE); (2) Mean Squared Error (MSE); (3) Coefficient of Determination ($R^2$), which measures the proportion of variance in dependent variable explained by the model. For generative tasks targeting molecular SMILES strings, four key evaluation metrics are considered: (1) Top-1 Accuracy, defined as the ratio of predicted SMILES strings that exactly match ground-truth ones. (2) Levenshtein Distance (LEV)~\cite{levenshtein1966binary}, measuring the minimum edits (insert, delete, substitute) to align two strings. (3) Tanimoto coefficient of molecular molar fingerprinting (MFP-TC) between predicted and ground-truth SMILES. (4) Invalidity Rate, representing the proportion of predicted SMILES strings unparsable as valid molecules by RDKit. Among these evaluation metrics, higher $R^2$, Top-1 accuracy, and MFP-TC are preferred, while lower MAE, MSE, LEV,
and invalidity rate are better.

\subsection{Implementation Details and Experimental Setup}
Mol-Tokenizer and Uni-Mol3 are both implemented in \texttt{Python 3.9} on 8 NVIDIA H100 GPUs (each with 81,920 MiB memory). Besides, RDKit (\texttt{version 2024.9.6}) served as the primary toolkit for molecular parsing and feature construction. All 3D molecular structures were generated using the ETKGD~\cite{riniker2015better} method and
optimized with the Merck Molecular Force Field (MMFF)~\cite{halgren1996merck} within the RDKit toolkit.

Mol-Tokenizer is initialized based on the pre-trained Uni-Mol2 model (\texttt{84M} version), with the following key architectural hyperparameters: encoder layer 12, FFN hidden dimension 748, pair hidden dimension 64, and number of attention heads 48. In addition, Mol-Tokenizer is trained with AdamW optimizer with weight decay 1e-4, learning rate 1e-4, batch size 256, training steps 100000, and warm-up steps 50000.

Uni-Mol3 adopts Text-to-Text Transfer Transformer (T5)~\cite{raffel2020exploring} as the backbone architecture with 8 layers of encoders, 8 layers of decoders, hidden dimension 768, and 8 attention heads. The training steps for molecular and reaction pre-training are 1,000,000 and 1,500,000, respectively, and other pre-training hyperparameters are the same: weight decay 1e-4, learning rate 1e-4 with polynomial scheduler for decaying learning rate, batch size 256, and warm-up steps 50000. For downstream fine-tuning, the dataset-specific hyperparameters for all datasets are summarized in Table.~\ref{tab:hyperparameter}.

\begin{table*}[!tbp]
\centering
\caption{Summary of dataset-specific hyperparameters, where there are two different learning rate (\texttt{lr}) schedules: ``fixed" denotes fixed learning rate during training, and ``polynomial" means decaying the learning rate using a polynomial function with power 1.0. For the Buchwald-Hartwig dataset, we use \texttt{max epoch} instead of \texttt{training steps} to control the number of model training iterations due to the limited dataset size.}
\vspace{0.5em}
\label{tab:hyperparameter}
\resizebox{1.\textwidth}{!}{
\begin{tabular}{lcccccc}
\toprule
\textbf{Dataset} & \textbf{training steps} & \textbf{batch size} & \textbf{lr} & \textbf{weight deacy} & \textbf{max epoch} & \textbf{lr scheduler} \\ \midrule
USPTO-MIT & 1,000,000 & 256 & 5e-4 & 1e-4 & - & polynomial \\
SMol-Reactions-FP & 2,000 & 256 & 5e-5 & 1e-4 & - & fixed \\
Pistachio-FP & 1,500,000 & 512 & 5e-4 & 1e-4 & - & polynomial \\ \midrule
USPTO-50k & 20,000 & 256 & 1e-4 & 1e-4 & - & polynomial \\
SMol-Reactions-RS & 10,000 & 256 & 1e-4 & 1e-4 & - & fixed \\
Pistachio-RS & 1,500,000 & 512 & 5e-4 & 1e-4 & - & polynomial \\ \midrule
USPTO-500-MT & 32,000 & 256 & 1e-4 & 1e-4 & - & fixed \\
USPTO-Condition & 20,000 & 256 & 1e-4 & 1e-4 & - & fixed \\
Pistachio-CG & 1,500,000 & 512 & 5e-4 & 1e-4 & - & polynomial \\ \midrule
Buchwald-Hartwig Test1 & - & 256 & 5e-4 & 1e-4 & 500 & polynomial \\
Buchwald-Hartwig Test2 & - & 512 & 1e-4 & 1e-4 & 500 & polynomial \\
Buchwald-Hartwig Test3 & - & 64 & 1e-4 & 0.0 & 100 & polynomial \\
Buchwald-Hartwig Test4 & - & 512 & 5e-4 & 5e-4 & 100 & polynomial \\ \bottomrule
\end{tabular}}

\end{table*}

\subsection{Results and Discussion}

\begin{table*}[!bp]
\begin{center}
\caption{Performance comparison for product prediction on the USPTO-MIT dataset, where reactant-condition separated and mixed are separately evaluated. The best and second results are marked as \textbf{bold} and \underline{underline}. (same for all the tables below)}
\label{tab:uspto_mit}
\resizebox{1.0\textwidth}{!}{
\begin{tabular}{l|cccc|cccc}
\toprule
\multirow{2}{*}{\textbf{Model}} & \multicolumn{4}{c}{\textbf{USPTO-MIT (Mixed)}} & \multicolumn{4}{c}{\textbf{USPTO-MIT (Seperated)}} \\ \cmidrule(r){2-5} \cmidrule(r){6-9} 
 & Top-1~(\%) & LEV & MFP-TC & Invalid~(\%) & Top-1~(\%) & LEV & MFP-TC & Invalid~(\%) \\ \midrule
Molformer \cite{schwaller2019molecular} & 88.3 & 0.543 & 0.971 & 0.32 & 89.0 & 0.445 & 0.975 & 0.26 \\
Chemformer \cite{irwin2022chemformer} & 88.6 & \underline{0.514} & \underline{0.976} & 0.25 & 89.8 & 0.428 & 0.979 & 0.17 \\
T5Chem \cite{lu2022unified} & \underline{88.9} & 0.527 & 0.974 & \underline{0.20} & \underline{90.2} & \underline{0.414} & \underline{0.981} & \textbf{0.10} \\ \midrule
\rowcolor{gray!10} Uni-Mol3 (ours) & \textbf{89.6} & \textbf{0.485} & \textbf{0.979} & \textbf{0.15} & \textbf{90.8} & \textbf{0.387} & \textbf{0.983} & \underline{0.15} \\ \bottomrule
\end{tabular}}

\end{center}
\end{table*} 

\subsubsection{Task 1: (Forward) Product Prediction}
We use three datasets to fine-tune Uni-Mol3 for benchmarking product prediction: USPTO-MIT, SMol-Reaction-FP, and Pistachio-FP. For USPTO-MIT and Pistachio-FP, we explore two different input settings: reactant-condition separated and mixed. The separated setting explicitly separates reactants and reaction conditions in the input, while the mixed setting simulates real-world scenarios by combining them—a setting inherently adopted by the SMol-Reaction-FP dataset. For the fine-tuning strategy, we first fine-tune the pre-trained Uni-Mol3 on the large-scale Pistachio-FP dataset, based on which we further fine-tune the model on USPTO-MIT. To avoid information leakage from Pistachio-FP, we directly fine-tune the pre-trained model on SMol-Reaction-FP, which emphasizes train-test data discrepancy. Four different evaluation metrics are used: (1) Top-1 accuracy; (2) Levenshtein Distance (LEV); (3) Tanimoto coefficient of molecular molar fingerprinting (MFP-TC); and (4) molecular invalidity rate by RDKit parsing. Results for the three datasets are reported in Tables.~\ref{tab:uspto_mit},\ref{tab:pistachio_fp},\ref{tab:smol_reaction_fp}, demonstrating that Uni-Mol3 outperforms existing baselines significantly—particularly in the LEV and invalidity rate metrics. Notably, we observe that Uni-Mol3 excels in both separated and mixed input settings, with no significant degradation in Top-1 accuracy or MFP-TC under the more challenging mixed setting.

\begin{table*}[!tbp]
\begin{center}
\caption{Performance comparison for product prediction on the Pistachio-FP dataset, where reactant-condition separated and mixed are separately evaluated.}
\vspace{0.5em}
\label{tab:pistachio_fp}
\resizebox{1.0\textwidth}{!}{
\begin{tabular}{l|cccc|cccc}
\toprule
\multirow{2}{*}{\textbf{Model}} & \multicolumn{4}{c}{\textbf{Pistachio-FP (Mixed)}} & \multicolumn{4}{c}{\textbf{Pistachio-FP (Seperated)}} \\ \cmidrule(r){2-5} \cmidrule(r){6-9} 
 & Top-1 (\%) & LEV & MFP-TC & Invalid (\%) & Top-1 (\%) & LEV & MFP-TC & Invalid (\%) \\ \midrule
Molformer \cite{schwaller2019molecular} & 88.3 & 0.637 & 0.977 & 0.36 & 90.3 & 0.529 & 0.982 & 0.33 \\
Chemformer \cite{irwin2022chemformer} & 90.3 & 0.575 & 0.980 & 0.18 & 91.8 & 0.447 & 0.985 & \underline{0.14} \\
T5Chem \cite{lu2022unified} & \underline{90.9} & \underline{0.560} & \underline{0.982} & \underline{0.15} & \underline{92.1} & \underline{0.428} & \underline{0.986} & 0.18 \\ \midrule
\rowcolor{gray!10}Uni-Mol3 (ours) & \textbf{91.7} & \textbf{0.462} & \textbf{0.985} & \textbf{0.09} & \textbf{93.0} & \textbf{0.374} & \textbf{0.988} & \textbf{0.07} \\ \bottomrule
\end{tabular}}

\end{center}
\end{table*} 
\begin{table*}[!tbp]
\begin{center}
\vspace{-1em}
\caption{Results for product prediction on the SMol-Reactions-FP dataset.}
\label{tab:smol_reaction_fp}
\resizebox{0.65\textwidth}{!}{
\begin{tabular}{l|cccc}
\toprule
\multirow{2}{*}{\textbf{Model}} & \multicolumn{4}{c}{\textbf{SMol-Reactions-FP}} \\ \cmidrule(r){2-5}
 & Top-1 (\%) & LEV & MFP-TC & Invalid (\%) \\ \midrule
Molformer \cite{schwaller2019molecular} & 32.8 & 10.314 & 0.646 & 4.54 \\
Chemformer \cite{irwin2022chemformer} & 36.9 & \underline{7.849} & 0.718 & 0.72 \\
T5Chem \cite{lu2022unified} & \underline{37.2} & 8.030 & \textbf{0.735} & \underline{0.69} \\
PRESTO \cite{cao2024presto} & 35.4 & 9.582 & 0.685 & 1.65 \\ \midrule
\rowcolor{gray!10}Uni-Mol3 (ours) & \textbf{38.7} & \textbf{7.014} & \underline{0.732} & \textbf{0.64} \\ \bottomrule
\end{tabular}}

\end{center}
\end{table*} 

\subsubsection{Task 2: Retrosynthetic Prediction}
We conduct comparative evaluations of various baselines for the retrosynthetic prediction task across three datasets: USPTO-50k, Pistachio-RS, and SMol-Reactions-RS. Following the product prediction paradigm, we first perform large-scale fine-tuning on the Pistachio-RS dataset, and then further fine-tune the model on the USPTO-50k dataset. The pre-trained Uni-Mol3 is directly fine-tuned on the SMol-Reactions-RS dataset to prevent information leakage from Pistachio-RS. We report the results in Tables.~\ref{tab:retrosynthetic},~\ref{tab:smol_reaction_rs} using Top-1 accuracy, LEV, MFP-TC, and invalidity as metrics. Uni-Mol3 demonstrates remarkable adaptability in the retrosynthetic prediction task, particularly excelling on the SMol-Reactions-RS dataset. This experimental comparison not only highlights Uni-Mol3’s proficiency in learning reaction inverse mapping but also underscores its robustness in scenarios where substantial data distribution gaps exist.

\begin{table*}[!tbp]
\begin{center}
\caption{Retrosynthesis results on the Pistachio-RS and USPTO-50k datasets.}
\label{tab:retrosynthetic}
\resizebox{1.0\textwidth}{!}{
\begin{tabular}{l|cccc|cccc}
\toprule
\multirow{2}{*}{\textbf{Model}} & \multicolumn{4}{c}{\textbf{Pistachio-RS}} & \multicolumn{4}{c}{\textbf{USPTO-50k}} \\ \cmidrule(r){2-5} \cmidrule(r){6-9} 
 & Top-1 (\%) & LEV & MFP-TC & Invalid (\%) & Top-1 (\%) & LEV & MFP-TC & Invalid (\%) \\ \midrule
Molformer \cite{schwaller2019molecular} & 72.6 & 2.554 & 0.953 & 0.56 & 42.6 & 4.486 & 0.911 & 0.79 \\
Chemformer \cite{irwin2022chemformer} & 74.6 & 2.374 & 0.957 & \underline{0.24} & \textbf{52.3} & 4.218 & 0.908 & \underline{0.16} \\
T5Chem \cite{lu2022unified} & \underline{75.2} & \underline{2.247} & \underline{0.959} & 0.32 & 46.2 & \underline{3.959} & \underline{0.916} & 0.32 \\ \midrule
\rowcolor{gray!10}Uni-Mol3 (ours) & \textbf{76.9} & \textbf{2.145} & \textbf{0.963} & \textbf{0.18} & \underline{49.0} & \textbf{3.653} & \textbf{0.924} & \textbf{0.06} \\ \bottomrule
\end{tabular}}

\end{center}
\end{table*} 
\begin{table*}[!tbp]
\begin{center}
\vspace{-1em}
\caption{Results for retrosynthetic prediction on the SMol-Reactions-RS dataset.}
\label{tab:smol_reaction_rs}
\resizebox{0.65\textwidth}{!}{
\begin{tabular}{l|cccc}
\toprule
\multirow{2}{*}{\textbf{Model}} & \multicolumn{4}{c}{\textbf{SMol-Reactions-RS}} \\ \cmidrule(r){2-5}
 & Top-1 (\%) & LEV & MFP-TC & Invalid (\%) \\ \midrule
Molformer \cite{schwaller2019molecular} & 23.9 & 15.382 & 0.659 & 2.73 \\
Chemformer \cite{irwin2022chemformer} & 26.5 & 12.017 & 0.714 & \underline{0.24} \\
T5Chem \cite{lu2022unified} & \underline{28.0} & \underline{10.593} & \underline{0.758} & 0.26 \\
PRESTO \cite{cao2024presto} & 27.7 & 11.229 & 0.745 & 1.15 \\ \midrule
\rowcolor{gray!10}Uni-Mol3 (ours) & \textbf{29.1} & \textbf{9.933} & \textbf{0.786} & \textbf{0.20} \\ \bottomrule
\end{tabular}}
\end{center}
\end{table*}

\subsubsection{Task 3: Condition Generation}
In previous studies, condition prediction approaches typically defined a fixed number of catalysts, solvents, and reagents explicitly, formulating the task as multi-label classification. In contrast, this work focuses on condition generation—a more realistic and challenging scenario that does not pre-specify the categories or quantities of condition molecules. Three datasets, USPTO-500-MT, USPTO-Condition, and Pistachio-CG, and four metrics, Top-1 accuracy, LEV, MFP-TC, and invalidity, are used for benchmarking condition generation. In a similar way, we first fine-tune the pre-trained Uni-Mol3 on the large-scale Pistachio-CG dataset, and then further fine-tune the model with two small-scale datasets, USPTO-500-MT and USPTO-Condition. As demonstrated in Tables.~\ref{tab:pistachio_cg} and ~\ref{tab:condition}, Uni-Mol3 outperforms all baseline models across all four evaluation metrics, highlighting its comprehensive superiority in handling open-ended condition generation without pre-defined molecular constraints.

\begin{table*}[!htbp]
\begin{center}
\caption{Results for condition generation on the Pistachio-CG dataset.}
\vspace{0.5em}
\label{tab:pistachio_cg}
\resizebox{0.65\textwidth}{!}{
\begin{tabular}{l|cccc}
\toprule
\multirow{2}{*}{\textbf{Model}} & \multicolumn{4}{c}{\textbf{Pistachio-CG}} \\ \cmidrule(r){2-5}
 & Top-1 (\%) & LEV & MFP-TC & Invalid (\%) \\ \midrule
Molformer \cite{schwaller2019molecular} & 40.4 & 7.272 & 0.810 & 0.11 \\
Chemformer \cite{irwin2022chemformer} & 42.1 & 6.947 & 0.819 & \underline{0.04} \\
T5Chem \cite{lu2022unified} & \underline{43.3} & \underline{6.705} & \underline{0.823} & 0.06 \\ \midrule
\rowcolor{gray!10}Uni-Mol3 (ours) & \textbf{44.4} & \textbf{6.482} & \textbf{0.827} & \textbf{0.03} \\ \bottomrule
\end{tabular}}

\end{center}
\end{table*} 
\begin{table*}[!htbp]
\begin{center}
\vspace{-1em}
\caption{Results for condition generation on USPTO-500-MT and USPTO-Condition.}
\label{tab:condition}
\resizebox{1.0\textwidth}{!}{
\begin{tabular}{l|cccc|cccc}
\toprule
\multirow{2}{*}{\textbf{Model}} & \multicolumn{4}{c}{\textbf{USPTO-500-MT}} & \multicolumn{4}{c}{\textbf{USPTO-Condition}} \\ \cmidrule(r){2-5} \cmidrule(r){6-9} 
 & Top-1 (\%) & LEV & MFP-TC & Invalid (\%) & Top-1 (\%) & LEV & MFP-TC & Invalid (\%) \\ \midrule
Molformer \cite{schwaller2019molecular} & 19.9 & 9.773 & 0.694 & 0.068 & 25.6 & 5.439 & 0.739 & 0.031 \\
Chemformer \cite{irwin2022chemformer} & 24.1 & 8.655 & 0.707 & 0.027 & 29.3 & 5.215 & 0.744 & 0.012 \\
T5Chem \cite{lu2022unified} & \textbf{24.9} & \underline{8.541} & \underline{0.712} & \underline{0.012} & \underline{29.8} & \textbf{5.087} & \underline{0.747} & \underline{0.004} \\ \midrule
\rowcolor{gray!10}Uni-Mol3 (ours) & \underline{24.5} & \textbf{8.523} & \textbf{0.715} & \textbf{0.007} & \textbf{30.5} & \underline{5.157} & \textbf{0.748} & \textbf{0.001} \\ \bottomrule
\end{tabular}}

\end{center}
\end{table*}

\subsubsection{Task 4: Reaction Yield Prediction}
We use the Buchwald-Hartwig dataset to evaluate Uni-Mol3’s performance in reaction yield prediction. The Buchwald-Hartwig dataset consists of high-throughput C-N coupling data and includes four different out-of-sample test sets that contain reaction additives not present in the training set. Evaluation metrics—MAE, RMSE, and R$^2$—are used to compare models across these test sets, with results summarized in Table.~\ref{tab:yield}. It can be found that Uni-Mol3 demonstrates overall superiority over baseline models, with the top performance on 11 out of 12 metrics. Notably, Uni-Mol3 outperforms T5Chem by 28.0\%, 18.2\%, and 7.2\% on MAE, RMSE, and R$^2$ on the Test1 set, respectively. For the Test4 set, while Uni-Mol3 ranks second to T5Chem in R$^2$, it leads in both MAE and RMSE. These results across all four test sets highlight Uni-Mol3's excellent generalization capability to out-of-distribution samples, a capability largely attributed to the large-scale reaction pre-training proposed in this paper.

\begin{table*}[!htbp]
\begin{center}
\caption{Results for yield prediction on 4 test sets of the Buchwald-Hartwig dataset.}
\label{tab:yield}
\resizebox{0.8\textwidth}{!}{
\begin{tabular}{l|ccc|ccc}
\toprule
\multirow{2}{*}{\textbf{Model}} & \multicolumn{3}{c}{\textbf{Test1}} & \multicolumn{3}{c}{\textbf{Test2}} \\ \cmidrule(r){2-4} \cmidrule(r){5-7}
 & MAE $\downarrow$ & RMSE $\downarrow$ & $R^2$ $\uparrow$ & MAE $\downarrow$ & RMSE $\downarrow$ & $R^2$ $\uparrow$ \\ \midrule
DRFP & 8.224 & 12.048 & 0.810 & 7.906 & 11.749 & 0.828 \\
Chemprop & 8.531 & 12.406 & 0.798 & 9.444 & 12.710 & 0.780 \\
YieldBert & 6.705 & 10.849 & 0.838 & 7.457 & 10.631 & 0.842 \\
T5Chem & \underline{8.145} & \underline{11.837} & \underline{0.815} & \underline{6.075} & \underline{8.784} & \underline{0.895} \\
\rowcolor{gray!10}Uni-Mol3 (ours) & \textbf{5.867} & \textbf{9.680} & \textbf{0.874} & \textbf{5.420} & \textbf{8.170} & \textbf{0.909} \\ \midrule
\multirow{2}{*}{\textbf{Model}} & \multicolumn{3}{c}{\textbf{Test1}} & \multicolumn{3}{c}{\textbf{Test2}} \\ \cmidrule(r){2-4} \cmidrule(r){5-7}
 & MAE $\downarrow$ & RMSE $\downarrow$ & $R^2$ $\uparrow$ & MAE $\downarrow$ & RMSE $\downarrow$ & $R^2$ $\uparrow$ \\ \midrule
DRFP & 9.525 & 14.880 & 0.719 & 13.240 & 19.037 & 0.496 \\
Chemprop & 10.340 & 15.280 & 0.708 & 15.783 & 20.155 & 0.429 \\
YieldBert & 9.109 & 14.136 & 0.746 & 13.045 & \underline{18.639} & 0.503 \\
T5Chem & \underline{8.977} & \underline{13.892} & \underline{0.765} & \underline{12.952} & 18.711 & \textbf{0.610} \\
\rowcolor{gray!10}Uni-Mol3 (ours) & \textbf{8.856} & \textbf{13.506} & \textbf{0.769} & \textbf{12.740} & \textbf{18.245} & \underline{0.525} \\ \bottomrule
\end{tabular}}

\end{center}
\end{table*} 

\subsubsection{Ablation Study and Analysis}
\begin{figure}[h]
\centering
    \includegraphics[width=0.75\textwidth]{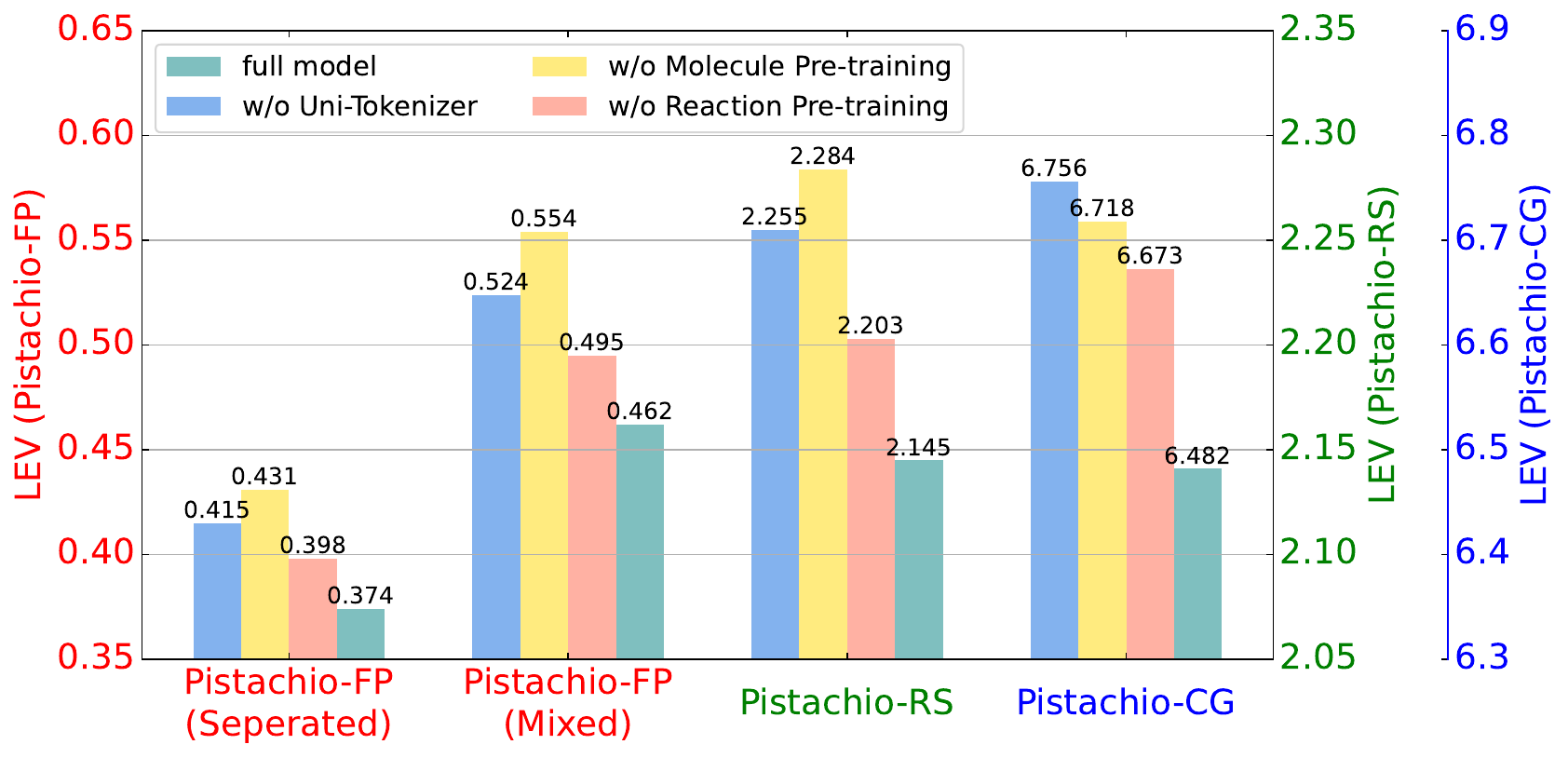}
    \caption{Ablation study on Uni-Tokenizer, molecular pre-training, and reaction pre-training on three Pistachio datasets with the Levenshtein Distance (LEV) as a metric.}
    \label{fig:ablation}
\end{figure}

To better evaluate the important roles played by the key modules in Uni-Mol3, we conduct a more in-depth analysis of the first three stages in the hierarchical pipeline in Figure.~\ref{fig:framework}, including Uni-Tokenizer, molecular pre-training, and reaction pre-training. We report in Figure.~\ref{fig:ablation} the ablation study on three Pistachio datasets for three tasks, where the product prediction task has two settings. We use the Levenshtein Distance (LEV) as a metric because it better reflects how close the generated results are to the ground-truth ones. For the ``w/o Uni-Tokenizer" set of experiments, we directly use SMILES strings as inputs. Key observations from Figure.~\ref{fig:ablation} reveal: (1) Molecular pre-training plays the most important role as it captures the fundamental grammatical rules of molecules. This capability ensures valid single-molecular generation, which forms the basis for multi-molecular reaction predictions. (2) The Uni-Tokenizer module enhances molecular representations with 3D structure-awareness, demonstrating huge performance gains across all reaction tasks on three Pistachio datasets, especially for the task of condition generation. (3) Reaction pre-training learns the fundamental rules of chemical reactions. While effective for all tasks, its performance improvements are slightly less pronounced than the other modules. This is largely attributed to overlapping knowledge between reaction pre-training and subsequent fine-tuning.

\subsubsection{Single-task \textit{vs.} Multi-task Evaluation}
We systematically compare the performance of T5Chem and Uni-Mol3 in single-task and multi-task prediction scenarios across three chemical reaction tasks—product prediction (separated setup), retrosynthetic prediction, and condition generation—using three Pistachio datasets. For single-task prediction, we directly fine-tune the pre-trained Uni-Mol3 with task-specific data and evaluate it on the corresponding test set. In contrast, for multi-task prediction, we mix the training data of all three tasks and sample randomly to create batches containing mixed-task data, enabling fine-tuning across multiple tasks. The fine-tuned model is then separately evaluated on the test set of each task. Three important observations are made from Table.~\ref{tab:multi_task}: (1) Enhanced molecular validity: Multi-task mixed fine-tuning significantly improves the validity of generated molecules for both T5Chem and Uni-Mol3. (2) Performance trade-off: Multi-task fine-tuning inherently involves a trade-off between tasks, resulting in performance drops for simpler tasks. For instance, both models show poor performance in product prediction compared to single-task fine-tuning, though Uni-Mol3's decline is less pronounced. (3) Model-specific benefits: T5Chem underperforms in multi-task prediction across all tasks, as measured by Top-1 accuracy and LEV metrics. Conversely, Uni-Mol3 benefits substantially from multi-task fine-tuning, particularly in retrosynthetic prediction and condition generation, where it outperforms single-task fine-tuning. Specifically, in the multi-task setting, Uni-Mol3 achieves a 3.5\% and 3.2\% increase in Top-1 accuracy for retrosynthesis and condition generation, respectively.

\begin{table*}[!tbp]
\begin{center}
\caption{Performance comparison of T5Chem and Uni-Mol3 in single-task and multi-task fine-tuning across three reaction tasks on the Pistachio-FP, Pistachio-RS, and Pistachio-CG datasets. The performance gains and drops are marked in {\color[rgb]{0.4, 0.71, 0.376}green} and {\color[rgb]{0.8, 0.227, 0.141}red}.}
\vspace{0.5em}
\label{tab:multi_task}
\resizebox{1.\textwidth}{!}{
\begin{tabular}{lccccccccc}
\toprule
\multirow{2}{*}{\textbf{Model}} & \multicolumn{3}{c}{\textbf{Product Prediction}} & \multicolumn{3}{c}{\textbf{Retrosynthetic Prediction}} & \multicolumn{3}{c}{\textbf{Condition Generation}} \\ \cmidrule(r){2-4} \cmidrule(r){5-7} \cmidrule(r){8-10} 
 & Top-1 (\%) & LEV & Invalid (\%) & Top-1 (\%) & LEV & Invalid (\%) & Top-1 (\%) & LEV & Invalid (\%) \\ \midrule
T5Chem (Single-task) & 92.1 & 0.428 & 0.18 & 75.2 & 2.247 & 0.32 & 43.3 & 6.705 & 0.06 \\
T5Chem (Multi-task) & 90.4 & 0.524 & 0.16 & 74.5 & 2.278 & 0.14 & 41.1 & 7.060 & 0.04 \\
\rowcolor{gray!10}$\Delta_{\text{T5Chem}}$ & {\color[rgb]{0.8, 0.227, 0.141}-1.7} & {\color[rgb]{0.8, 0.227, 0.141}+0.096}  & {\color[rgb]{0.4, 0.71, 0.376}-0.02} & {\color[rgb]{0.8, 0.227, 0.141}-0.7} & {\color[rgb]{0.8, 0.227, 0.141}+0.031} & {\color[rgb]{0.4, 0.71, 0.376}-0.18} & {\color[rgb]{0.8, 0.227, 0.141}-2.2} & {\color[rgb]{0.8, 0.227, 0.141}+0.355} & {\color[rgb]{0.4, 0.71, 0.376}-0.02} \\ \midrule
Uni-Mol3 (Single-task) & 93.0 & 0.374 & 0.07 & 76.9 & 2.145 & 0.18 & 44.4 & 6.482 & 0.03 \\
Uni-Mol3 (Multi-task) & 92.3 & 0.405 & 0.05 & 80.4 & 1.601 & 0.07 & 47.6 & 5.771 & 0.01 \\
\rowcolor{gray!10}$\Delta_{\text{Uni-Mol3}}$ & {\color[rgb]{0.8, 0.227, 0.141}-0.7} & {\color[rgb]{0.8, 0.227, 0.141}+0.031} & {\color[rgb]{0.4, 0.71, 0.376}-0.02} & {\color[rgb]{0.4, 0.71, 0.376}+3.5} & {\color[rgb]{0.4, 0.71, 0.376}-0.544} & {\color[rgb]{0.4, 0.71, 0.376}-0.11} & {\color[rgb]{0.4, 0.71, 0.376}+3.2} & {\color[rgb]{0.4, 0.71, 0.376}-0.711} & {\color[rgb]{0.4, 0.71, 0.376}-0.02} \\ \bottomrule
\end{tabular}}

\end{center}
\end{table*}

\begin{figure}[h]
\centering
    \includegraphics[width=0.7\textwidth]{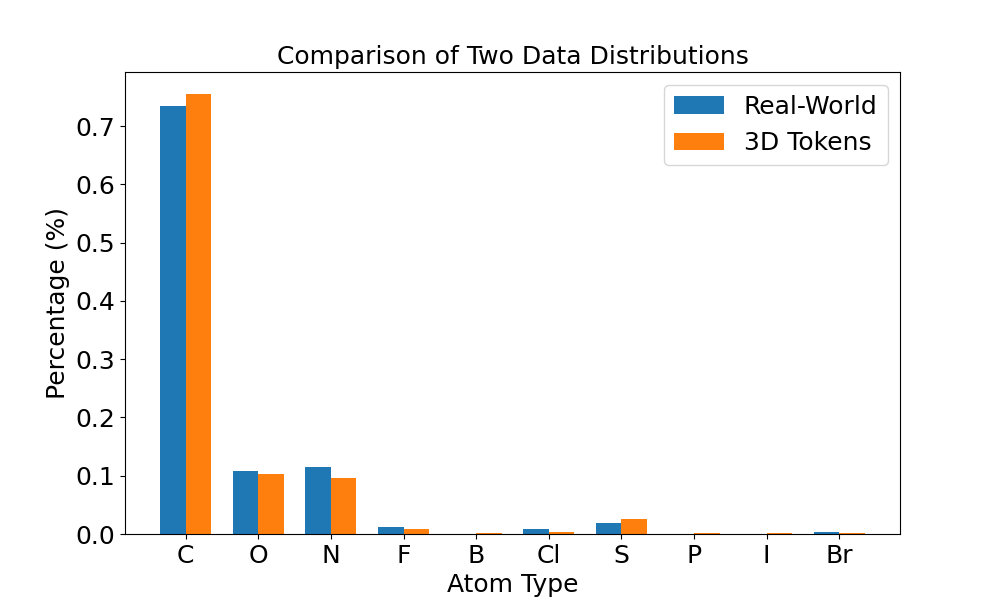}
    \caption{Distribution of atoms primarily encoded by each 3D token in Uni-Tokenizer, as well as the distribution of atoms in the real-world Pistachio-full dataset. The two distributions exhibit a high degree of similarity, i.e., atoms with higher natural abundance are allocated more 3D tokens in Uni-Tokenizer to encode their local environments.}
    \label{fig:distribution}
\end{figure}

\begin{figure}[b]
\centering
    \includegraphics[width=1.0\textwidth]{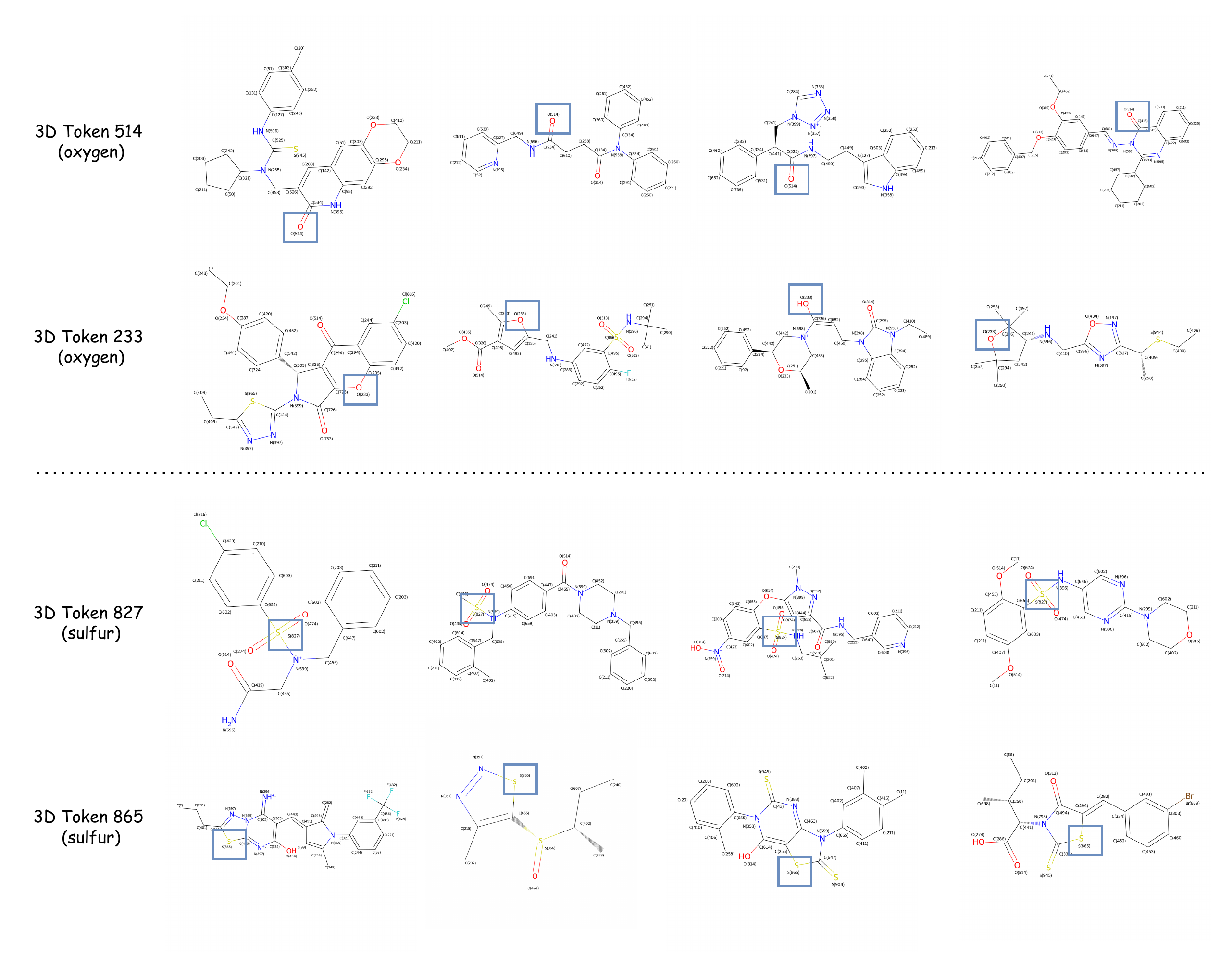}
    \caption{Visualization of local neighborhoods of 3D tokens mapped to the same atom, where the same atom can map to different 3D tokens due to variations in local contexts.}
    \label{fig:tokenization}
\end{figure}

\subsubsection{Visualization and analysis on Uni-Tokenize}
We present the real-world atomic distribution in the Pistachio-full dataset and the distribution of atoms encoded by each 3D token in Uni-Tokenizer in Figure.~\ref{fig:distribution}. Notably, the two distributions exhibit a high degree of similarity—atoms with higher natural abundance are allocated more 3D tokens in Uni-Tokenizer to encode their local environments. Further, we visualize the atoms encoded by distinct 3D tokens as well as their local neighborhoods in Figure.~\ref{fig:tokenization}. The results show that the same atom can map to different 3D tokens due to variations in local contexts, and different tokens encoding the same atom exhibit distinct local neighborhood patterns. For instance, 3D Token 514 primarily encodes oxygen atoms engaged in double bonds with adjacent atoms, while 3D Token 233 focuses on oxygen atoms connected via two single bonds. As another example, 3D Token 827 is specialized for sulfur atoms double-bonded to two oxygen atoms, whereas 3D Token 865 is connected to the others by a single bond in one ring. The visualizations in Figure.~\ref{fig:tokenization} demonstrate that Uni-Tokenizer learns tokens with excellent correspondence to atomic identities and their local environmental patterns, highlighting the model's capacity to capture fine-grained contextual details.

\subsubsection{Domain Shift Generalizability}
\begin{figure}[h]
\centering
    \includegraphics[width=0.8\textwidth]{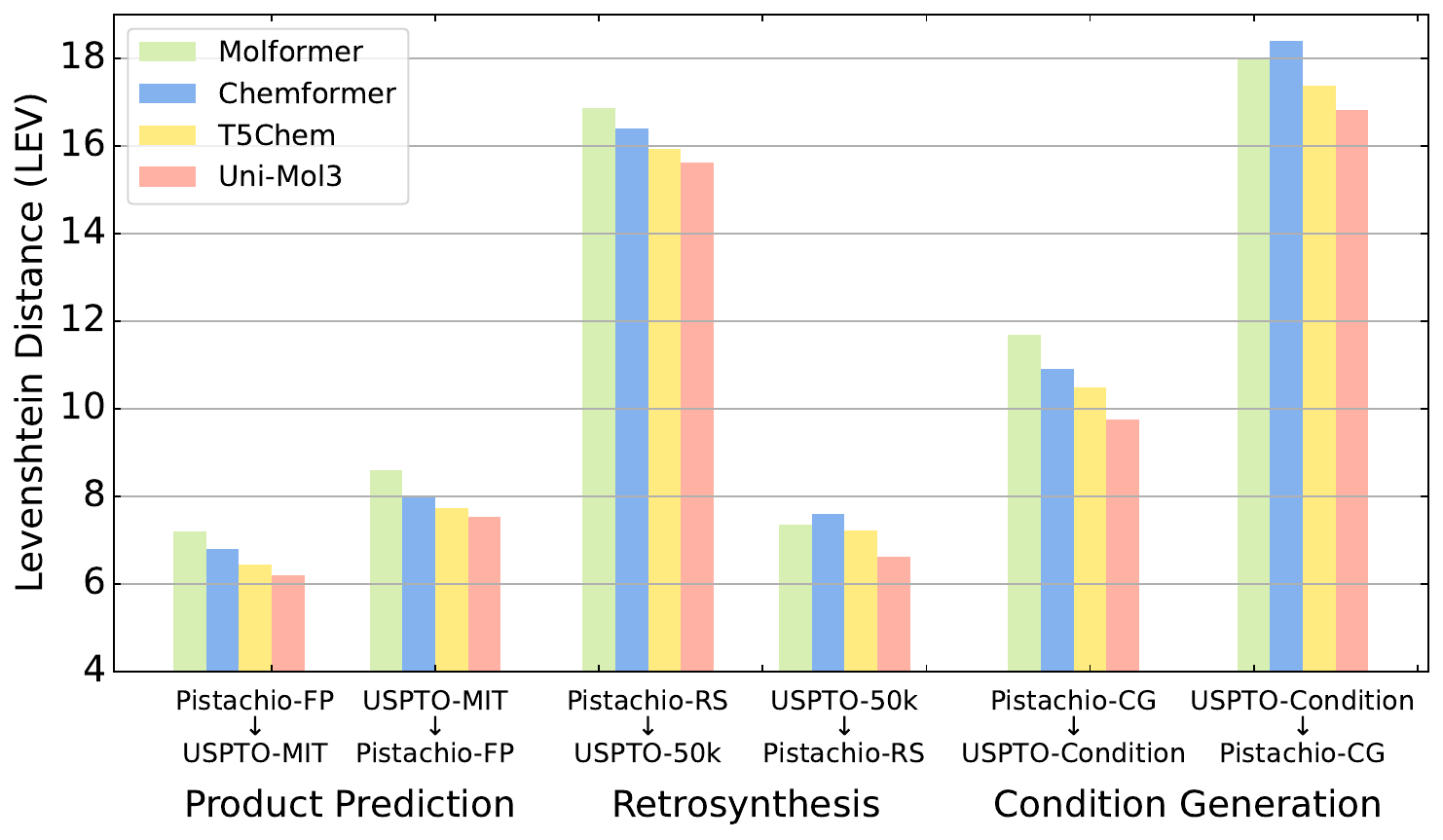}
    \caption{Cross-dataset generalization performance comparison of testing on unseen trainset-heterogenous test data for three chemical reaction tasks.}
    \label{fig:generalization}
\end{figure}

To further investigate the impact of domain shift on generalization, we conduct cross-dataset model transfer experiments by training models on one dataset and evaluating them on another. Focusing on three chemical reaction tasks—product prediction (in separated setting), retrosynthetic prediction, and reaction condition generation—we use Levenshtein Distance as the evaluation metric. Figure~\ref{fig:generalization} reports the performance of Molformer, Chemformer, T5Chem, and Uni-Mol3 under six distinct domain shift scenarios. Notably, Uni-Mol3 outperforms all baselines across all six settings, demonstrating particularly significant advantages in the condition generation task and when transferring from the large-scale Pistachio datasets to smaller-scale reaction datasets.

\begin{figure}[p]
\centering
    \includegraphics[width=1\textwidth]{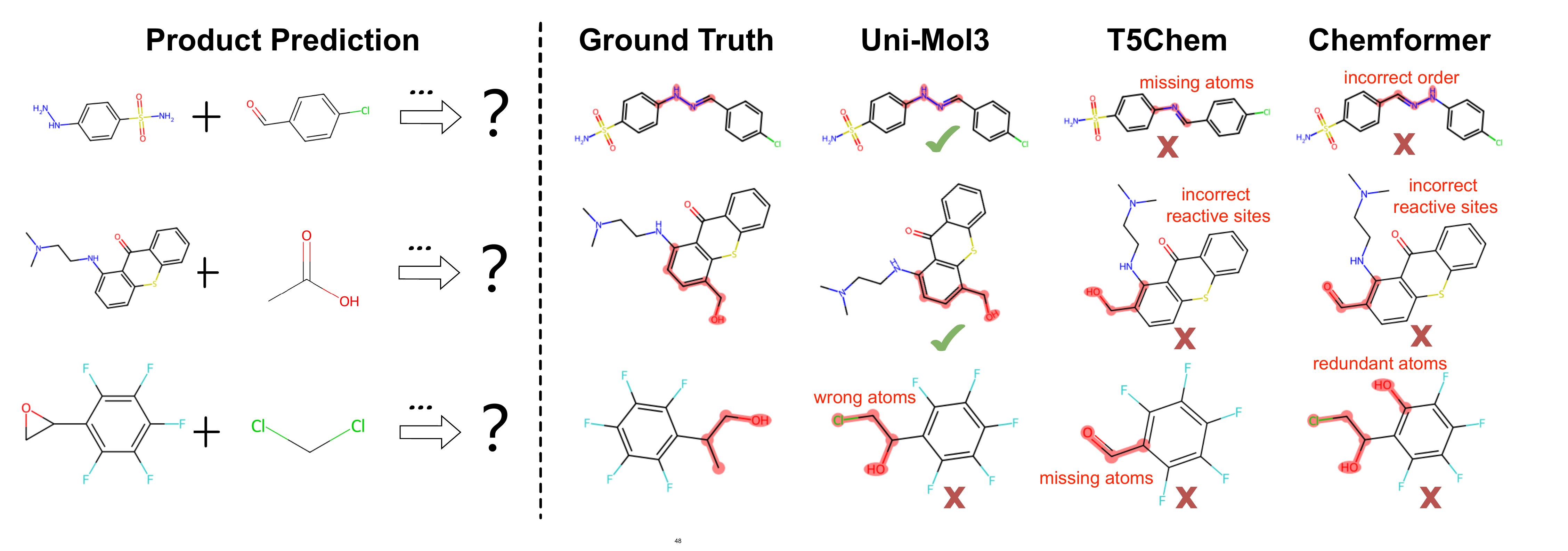}
    \includegraphics[width=1\textwidth]{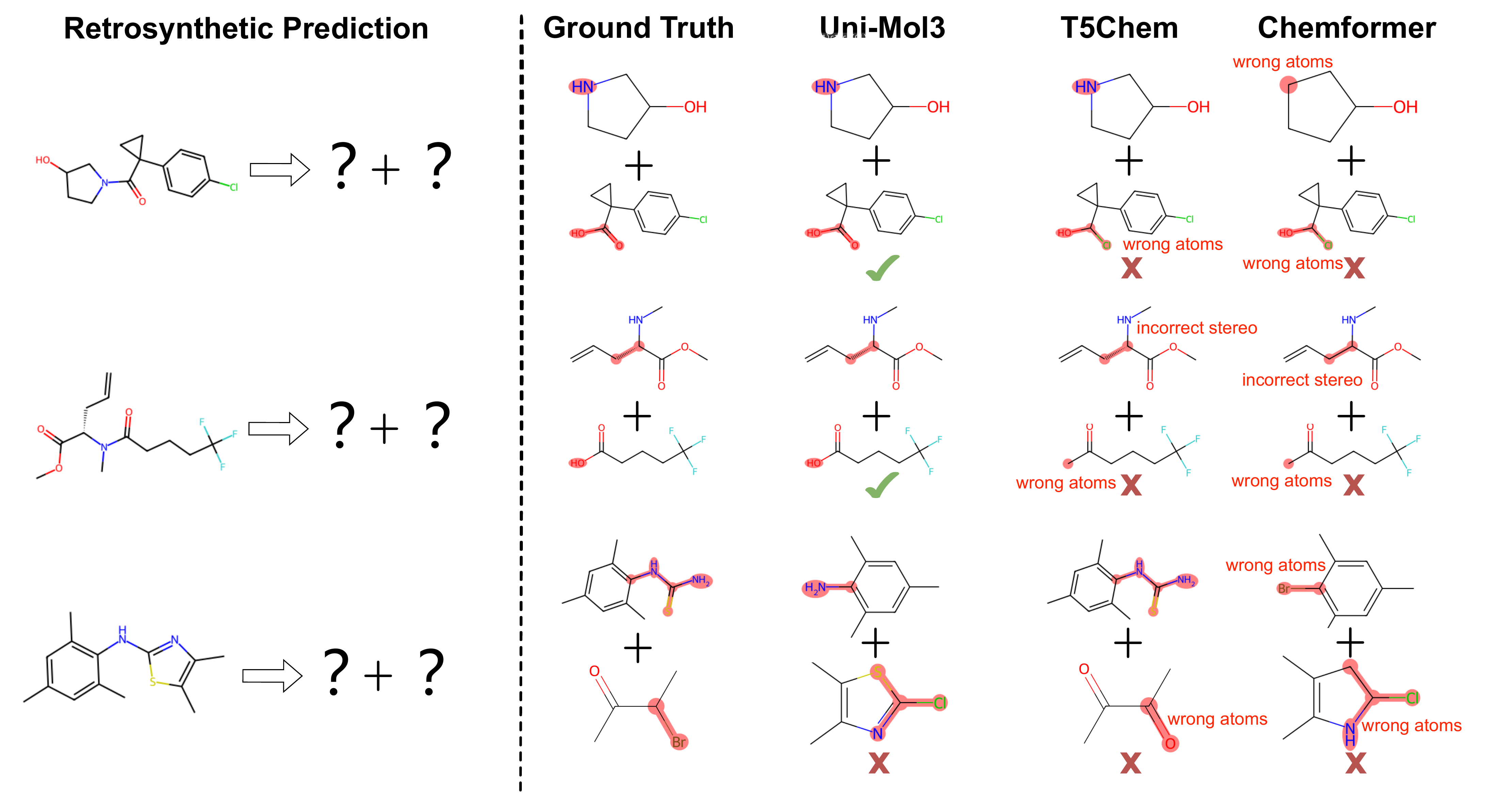}
    \includegraphics[width=1\textwidth]{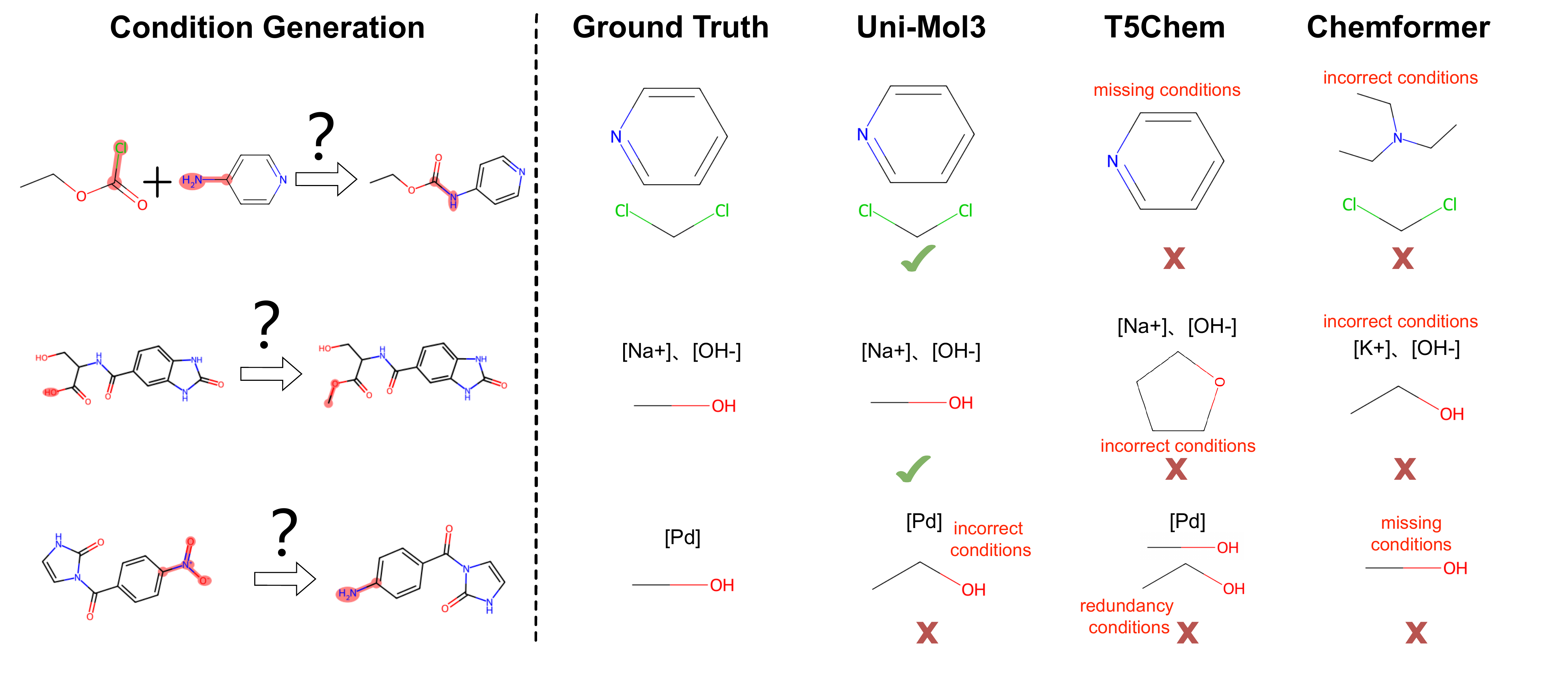}
    \caption{Case study of Chemformer, T5Chem, and Uni-Mol3 for three chemical reaction tasks, in which we highlight and annotate some common prediction errors.}
    \label{fig:vis_forward}
\end{figure}

\subsubsection{Case Study}
Figure.~\ref{fig:vis_forward} illustrates case studies of Chemformer, T5Chem, Uni-Mol3 as well as ground-truth ones for three tasks: product prediction, retrosynthetic prediction, and condition generation. We select several representative examples from three Pistachio datasets to demonstrate in which cases Uni-Mol3 works or fails, where we highlight and annotate prediction errors. Taking product prediction as an example, products generated by Chemformer and T5Chem may have issues such as missing atoms, incorrect atomic ordering, and incorrect reactive sites on the ring. In contrast, Uni-Mol3 shows far fewer such errors—though it occasionally generates wrong atoms, leading to slight deviations from the ground truth. For retrosynthetic prediction, Chemformer and T5Chem commonly generate reactants with wrong atoms, incorrect functional groups, or misformed bonds. Uni-Mol3 rarely commits these errors but may produce reactants that differ significantly from the ground truth. Notably, closer analysis reveals that these seemingly incorrect reactants often represent alternative retrosynthetic pathways, potentially under different reaction conditions. This suggests Uni-Mol3 possesses strong generative versatility for retrosynthesis, capable of offering multiple synthetic solutions. For the condition generation task, all three models struggle with generating correct reaction conditions, often omitting, adding, or misidentifying key conditions. For instance, Uni-Mol3 generates similar solvents, e.g., formaldehyde and ethanol.

\section{Conclusion}
This study introduces Uni-Mol3, a deep learning framework for multi-molecular organic reaction modeling. By integrating 3D structure-aware molecular tokenization (Mol-Tokenizer), hierarchical pre-training, and prompt-aware fine-tuning, the model achieves state-of-the-art performance across diverse reaction tasks. The multi-scale Mol-Tokenizer encodes 1D atomic features, 2D graph structures, and 3D spatial coordinates into discrete tokens, addressing the inherent limitations of SMILES in capturing spatial information. The two-tier pre-training strategy—first learning single-molecular grammatical rules, then capturing multi-molecular reaction principles—establishes a progressive learning paradigm from single- to multi-molecular systems. With prompt-adaptive fine-tuning, Uni-Mol3 enables seamless adaptation to product prediction, retrosynthetic prediction, reaction condition generation, and yield prediction with minimal architectural modifications. Extensive experiments on 10 datasets and 4 downstream tasks demonstrate Uni-Mol3's significant superiority over existing baselines. By unifying single and multi-molecular modeling, Uni-Mol3 defines a versatile framework for intelligent reactions, that promises to advance data-driven innovation in organic synthesis and accelerate its translation to industrial applications.

\section{Data and Code Availability}
The open-source implementation of Uni-Mol3 is available at \url{https://github.com/LirongWu/Uni-Mol3}, encompassing preprocessed data, preprocessing scripts, model weights, pre-training and fine-tuning pipelines, as well as evaluation protocols. For all publicly accessible datasets, we provide details on data size, usage instructions, and preprocessed data via the website. Due to licensing constraints, Pistachio-related datasets—including Pistachio-full, Pistachio-FP, Pistachio-RS, and Pistachio-CG—cannot be publicly distributed. However, we make the corresponding data preprocessing scripts available; researchers with official access licenses to the Pistachio datasets can utilize these scripts to generate the preprocessed data.

\bibliography{sn-bibliography}


\begin{thebibliography}{45}
\ifx \bisbn   \undefined \def \bisbn  #1{ISBN #1}\fi
\ifx \binits  \undefined \def \binits#1{#1}\fi
\ifx \bauthor  \undefined \def \bauthor#1{#1}\fi
\ifx \batitle  \undefined \def \batitle#1{#1}\fi
\ifx \bjtitle  \undefined \def \bjtitle#1{#1}\fi
\ifx \bvolume  \undefined \def \bvolume#1{\textbf{#1}}\fi
\ifx \byear  \undefined \def \byear#1{#1}\fi
\ifx \bissue  \undefined \def \bissue#1{#1}\fi
\ifx \bfpage  \undefined \def \bfpage#1{#1}\fi
\ifx \blpage  \undefined \def \blpage #1{#1}\fi
\ifx \burl  \undefined \def \burl#1{\textsf{#1}}\fi
\ifx \doiurl  \undefined \def \doiurl#1{\url{https://doi.org/#1}}\fi
\ifx \betal  \undefined \def \betal{\textit{et al.}}\fi
\ifx \binstitute  \undefined \def \binstitute#1{#1}\fi
\ifx \binstitutionaled  \undefined \def \binstitutionaled#1{#1}\fi
\ifx \bctitle  \undefined \def \bctitle#1{#1}\fi
\ifx \beditor  \undefined \def \beditor#1{#1}\fi
\ifx \bpublisher  \undefined \def \bpublisher#1{#1}\fi
\ifx \bbtitle  \undefined \def \bbtitle#1{#1}\fi
\ifx \bedition  \undefined \def \bedition#1{#1}\fi
\ifx \bseriesno  \undefined \def \bseriesno#1{#1}\fi
\ifx \blocation  \undefined \def \blocation#1{#1}\fi
\ifx \bsertitle  \undefined \def \bsertitle#1{#1}\fi
\ifx \bsnm \undefined \def \bsnm#1{#1}\fi
\ifx \bsuffix \undefined \def \bsuffix#1{#1}\fi
\ifx \bparticle \undefined \def \bparticle#1{#1}\fi
\ifx \barticle \undefined \def \barticle#1{#1}\fi
\bibcommenthead
\ifx \bconfdate \undefined \def \bconfdate #1{#1}\fi
\ifx \botherref \undefined \def \botherref #1{#1}\fi
\ifx \url \undefined \def \url#1{\textsf{#1}}\fi
\ifx \bchapter \undefined \def \bchapter#1{#1}\fi
\ifx \bbook \undefined \def \bbook#1{#1}\fi
\ifx \bcomment \undefined \def \bcomment#1{#1}\fi
\ifx \oauthor \undefined \def \oauthor#1{#1}\fi
\ifx \citeauthoryear \undefined \def \citeauthoryear#1{#1}\fi
\ifx \endbibitem  \undefined \def \endbibitem {}\fi
\ifx \bconflocation  \undefined \def \bconflocation#1{#1}\fi
\ifx \arxivurl  \undefined \def \arxivurl#1{\textsf{#1}}\fi
\csname PreBibitemsHook\endcsname

\bibitem[\protect\citeauthoryear{Zhou et~al.}{2023}]{zhou2023uni}
\begin{botherref}
\oauthor{\bsnm{Zhou}, \binits{G.}},
\oauthor{\bsnm{Gao}, \binits{Z.}},
\oauthor{\bsnm{Ding}, \binits{Q.}},
\oauthor{\bsnm{Zheng}, \binits{H.}},
\oauthor{\bsnm{Xu}, \binits{H.}},
\oauthor{\bsnm{Wei}, \binits{Z.}},
\oauthor{\bsnm{Zhang}, \binits{L.}},
\oauthor{\bsnm{Ke}, \binits{G.}}:
Uni-mol: A universal 3d molecular representation learning framework
(2023)
\end{botherref}
\endbibitem

\bibitem[\protect\citeauthoryear{Ji et~al.}{2024}]{ji2024uni}
\begin{botherref}
\oauthor{\bsnm{Ji}, \binits{X.}},
\oauthor{\bsnm{Wang}, \binits{Z.}},
\oauthor{\bsnm{Gao}, \binits{Z.}},
\oauthor{\bsnm{Zheng}, \binits{H.}},
\oauthor{\bsnm{Zhang}, \binits{L.}},
\oauthor{\bsnm{Ke}, \binits{G.}}, et al.:
Uni-mol2: Exploring molecular pretraining model at scale.
arXiv preprint arXiv:2406.14969
(2024)
\end{botherref}
\endbibitem

\bibitem[\protect\citeauthoryear{Chu et~al.}{2021}]{chu2021desulfonylation}
\begin{barticle}
\bauthor{\bsnm{Chu}, \binits{X.-Q.}},
\bauthor{\bsnm{Ge}, \binits{D.}},
\bauthor{\bsnm{Cui}, \binits{Y.-Y.}},
\bauthor{\bsnm{Shen}, \binits{Z.-L.}},
\bauthor{\bsnm{Li}, \binits{C.-J.}}:
\batitle{Desulfonylation via radical process: recent developments in organic synthesis}.
\bjtitle{Chemical reviews}
\bvolume{121}(\bissue{20}),
\bfpage{12548}--\blpage{12680}
(\byear{2021})
\end{barticle}
\endbibitem

\bibitem[\protect\citeauthoryear{Ali et~al.}{2024}]{ali2024machine}
\begin{barticle}
\bauthor{\bsnm{Ali}, \binits{R.S.A.E.}},
\bauthor{\bsnm{Meng}, \binits{J.}},
\bauthor{\bsnm{Khan}, \binits{M.E.I.}},
\bauthor{\bsnm{Jiang}, \binits{X.}}:
\batitle{Machine learning advancements in organic synthesis: A focused exploration of artificial intelligence applications in chemistry}.
\bjtitle{Artificial Intelligence Chemistry}
\bvolume{2}(\bissue{1}),
\bfpage{100049}
(\byear{2024})
\end{barticle}
\endbibitem

\bibitem[\protect\citeauthoryear{Oliveira et~al.}{2022}]{oliveira2022machine}
\begin{barticle}
\bauthor{\bsnm{Oliveira}, \binits{J.C.}},
\bauthor{\bsnm{Frey}, \binits{J.}},
\bauthor{\bsnm{Zhang}, \binits{S.-Q.}},
\bauthor{\bsnm{Xu}, \binits{L.-C.}},
\bauthor{\bsnm{Li}, \binits{X.}},
\bauthor{\bsnm{Li}, \binits{S.-W.}},
\bauthor{\bsnm{Hong}, \binits{X.}},
\bauthor{\bsnm{Ackermann}, \binits{L.}}:
\batitle{When machine learning meets molecular synthesis}.
\bjtitle{Trends in Chemistry}
\bvolume{4}(\bissue{10}),
\bfpage{863}--\blpage{885}
(\byear{2022})
\end{barticle}
\endbibitem

\bibitem[\protect\citeauthoryear{Dong et~al.}{2022}]{dong2022deep}
\begin{barticle}
\bauthor{\bsnm{Dong}, \binits{J.}},
\bauthor{\bsnm{Zhao}, \binits{M.}},
\bauthor{\bsnm{Liu}, \binits{Y.}},
\bauthor{\bsnm{Su}, \binits{Y.}},
\bauthor{\bsnm{Zeng}, \binits{X.}}:
\batitle{Deep learning in retrosynthesis planning: datasets, models and tools}.
\bjtitle{Briefings in Bioinformatics}
\bvolume{23}(\bissue{1}),
\bfpage{391}
(\byear{2022})
\end{barticle}
\endbibitem

\bibitem[\protect\citeauthoryear{Chen et~al.}{2024}]{chen2024reaction}
\begin{barticle}
\bauthor{\bsnm{Chen}, \binits{S.}},
\bauthor{\bsnm{Noh}, \binits{J.}},
\bauthor{\bsnm{Jang}, \binits{J.}},
\bauthor{\bsnm{Kim}, \binits{S.}},
\bauthor{\bsnm{Gu}, \binits{G.H.}},
\bauthor{\bsnm{Jung}, \binits{Y.}}:
\batitle{Reaction templates: Bridging synthesis knowledge and artificial intelligence}.
\bjtitle{Accounts of Chemical Research}
\bvolume{57}(\bissue{14}),
\bfpage{1964}--\blpage{1972}
(\byear{2024})
\end{barticle}
\endbibitem

\bibitem[\protect\citeauthoryear{Seidl et~al.}{2022}]{seidl2022improving}
\begin{barticle}
\bauthor{\bsnm{Seidl}, \binits{P.}},
\bauthor{\bsnm{Renz}, \binits{P.}},
\bauthor{\bsnm{Dyubankova}, \binits{N.}},
\bauthor{\bsnm{Neves}, \binits{P.}},
\bauthor{\bsnm{Verhoeven}, \binits{J.}},
\bauthor{\bsnm{Wegner}, \binits{J.K.}},
\bauthor{\bsnm{Segler}, \binits{M.}},
\bauthor{\bsnm{Hochreiter}, \binits{S.}},
\bauthor{\bsnm{Klambauer}, \binits{G.}}:
\batitle{Improving few-and zero-shot reaction template prediction using modern hopfield networks}.
\bjtitle{Journal of chemical information and modeling}
\bvolume{62}(\bissue{9}),
\bfpage{2111}--\blpage{2120}
(\byear{2022})
\end{barticle}
\endbibitem

\bibitem[\protect\citeauthoryear{Wang et~al.}{2025}]{wang2025reacon}
\begin{barticle}
\bauthor{\bsnm{Wang}, \binits{Z.}},
\bauthor{\bsnm{Lin}, \binits{K.}},
\bauthor{\bsnm{Pei}, \binits{J.}},
\bauthor{\bsnm{Lai}, \binits{L.}}:
\batitle{Reacon: a template-and cluster-based framework for reaction condition prediction}.
\bjtitle{Chemical Science}
\bvolume{16}(\bissue{2}),
\bfpage{854}--\blpage{866}
(\byear{2025})
\end{barticle}
\endbibitem

\bibitem[\protect\citeauthoryear{Das et~al.}{2024}]{das2024advances}
\begin{barticle}
\bauthor{\bsnm{Das}, \binits{M.}},
\bauthor{\bsnm{Ghosh}, \binits{A.}},
\bauthor{\bsnm{Sunoj}, \binits{R.B.}}:
\batitle{Advances in machine learning with chemical language models in molecular property and reaction outcome predictions}.
\bjtitle{Journal of Computational Chemistry}
\bvolume{45}(\bissue{14}),
\bfpage{1160}--\blpage{1176}
(\byear{2024})
\end{barticle}
\endbibitem

\bibitem[\protect\citeauthoryear{Li et~al.}{2024}]{li2024research}
\begin{bchapter}
\bauthor{\bsnm{Li}, \binits{Y.}},
\bauthor{\bsnm{Zhao}, \binits{W.}},
\bauthor{\bsnm{Dang}, \binits{B.}},
\bauthor{\bsnm{Yan}, \binits{X.}},
\bauthor{\bsnm{Gao}, \binits{M.}},
\bauthor{\bsnm{Wang}, \binits{W.}},
\bauthor{\bsnm{Xiao}, \binits{M.}}:
\bctitle{Research on adverse drug reaction prediction model combining knowledge graph embedding and deep learning}.
In: \bbtitle{2024 4th International Conference on Machine Learning and Intelligent Systems Engineering (MLISE)},
pp. \bfpage{322}--\blpage{329}
(\byear{2024}).
\bcomment{IEEE}
\end{bchapter}
\endbibitem

\bibitem[\protect\citeauthoryear{Zhong et~al.}{2024}]{zhong2024recent}
\begin{barticle}
\bauthor{\bsnm{Zhong}, \binits{Z.}},
\bauthor{\bsnm{Song}, \binits{J.}},
\bauthor{\bsnm{Feng}, \binits{Z.}},
\bauthor{\bsnm{Liu}, \binits{T.}},
\bauthor{\bsnm{Jia}, \binits{L.}},
\bauthor{\bsnm{Yao}, \binits{S.}},
\bauthor{\bsnm{Hou}, \binits{T.}},
\bauthor{\bsnm{Song}, \binits{M.}}:
\batitle{Recent advances in deep learning for retrosynthesis}.
\bjtitle{Wiley Interdisciplinary Reviews: Computational Molecular Science}
\bvolume{14}(\bissue{1}),
\bfpage{1694}
(\byear{2024})
\end{barticle}
\endbibitem

\bibitem[\protect\citeauthoryear{Schwaller et~al.}{2019}]{schwaller2019molecular}
\begin{barticle}
\bauthor{\bsnm{Schwaller}, \binits{P.}},
\bauthor{\bsnm{Laino}, \binits{T.}},
\bauthor{\bsnm{Gaudin}, \binits{T.}},
\bauthor{\bsnm{Bolgar}, \binits{P.}},
\bauthor{\bsnm{Hunter}, \binits{C.A.}},
\bauthor{\bsnm{Bekas}, \binits{C.}},
\bauthor{\bsnm{Lee}, \binits{A.A.}}:
\batitle{Molecular transformer: a model for uncertainty-calibrated chemical reaction prediction}.
\bjtitle{ACS central science}
\bvolume{5}(\bissue{9}),
\bfpage{1572}--\blpage{1583}
(\byear{2019})
\end{barticle}
\endbibitem

\bibitem[\protect\citeauthoryear{Dai et~al.}{2019}]{dai2019retrosynthesis}
\begin{botherref}
\oauthor{\bsnm{Dai}, \binits{H.}},
\oauthor{\bsnm{Li}, \binits{C.}},
\oauthor{\bsnm{Coley}, \binits{C.}},
\oauthor{\bsnm{Dai}, \binits{B.}},
\oauthor{\bsnm{Song}, \binits{L.}}:
Retrosynthesis prediction with conditional graph logic network.
Advances in Neural Information Processing Systems
\textbf{32}
(2019)
\end{botherref}
\endbibitem

\bibitem[\protect\citeauthoryear{Lu and Zhang}{2022}]{lu2022unified}
\begin{barticle}
\bauthor{\bsnm{Lu}, \binits{J.}},
\bauthor{\bsnm{Zhang}, \binits{Y.}}:
\batitle{Unified deep learning model for multitask reaction predictions with explanation}.
\bjtitle{Journal of chemical information and modeling}
\bvolume{62}(\bissue{6}),
\bfpage{1376}--\blpage{1387}
(\byear{2022})
\end{barticle}
\endbibitem

\bibitem[\protect\citeauthoryear{Coley et~al.}{2017}]{coley2017prediction}
\begin{barticle}
\bauthor{\bsnm{Coley}, \binits{C.W.}},
\bauthor{\bsnm{Barzilay}, \binits{R.}},
\bauthor{\bsnm{Jaakkola}, \binits{T.S.}},
\bauthor{\bsnm{Green}, \binits{W.H.}},
\bauthor{\bsnm{Jensen}, \binits{K.F.}}:
\batitle{Prediction of organic reaction outcomes using machine learning}.
\bjtitle{ACS central science}
\bvolume{3}(\bissue{5}),
\bfpage{434}--\blpage{443}
(\byear{2017})
\end{barticle}
\endbibitem

\bibitem[\protect\citeauthoryear{Shi et~al.}{2020}]{shi2020graph}
\begin{bchapter}
\bauthor{\bsnm{Shi}, \binits{C.}},
\bauthor{\bsnm{Xu}, \binits{M.}},
\bauthor{\bsnm{Guo}, \binits{H.}},
\bauthor{\bsnm{Zhang}, \binits{M.}},
\bauthor{\bsnm{Tang}, \binits{J.}}:
\bctitle{A graph to graphs framework for retrosynthesis prediction}.
In: \bbtitle{International Conference on Machine Learning},
pp. \bfpage{8818}--\blpage{8827}
(\byear{2020}).
\bcomment{PMLR}
\end{bchapter}
\endbibitem

\bibitem[\protect\citeauthoryear{Ahneman et~al.}{2018}]{ahneman2018predicting}
\begin{barticle}
\bauthor{\bsnm{Ahneman}, \binits{D.T.}},
\bauthor{\bsnm{Estrada}, \binits{J.G.}},
\bauthor{\bsnm{Lin}, \binits{S.}},
\bauthor{\bsnm{Dreher}, \binits{S.D.}},
\bauthor{\bsnm{Doyle}, \binits{A.G.}}:
\batitle{Predicting reaction performance in c--n cross-coupling using machine learning}.
\bjtitle{Science}
\bvolume{360}(\bissue{6385}),
\bfpage{186}--\blpage{190}
(\byear{2018})
\end{barticle}
\endbibitem

\bibitem[\protect\citeauthoryear{Probst et~al.}{2022}]{probst2022reaction}
\begin{barticle}
\bauthor{\bsnm{Probst}, \binits{D.}},
\bauthor{\bsnm{Schwaller}, \binits{P.}},
\bauthor{\bsnm{Reymond}, \binits{J.-L.}}:
\batitle{Reaction classification and yield prediction using the differential reaction fingerprint drfp}.
\bjtitle{Digital discovery}
\bvolume{1}(\bissue{2}),
\bfpage{91}--\blpage{97}
(\byear{2022})
\end{barticle}
\endbibitem

\bibitem[\protect\citeauthoryear{Schneider et~al.}{2015}]{schneider2015development}
\begin{barticle}
\bauthor{\bsnm{Schneider}, \binits{N.}},
\bauthor{\bsnm{Lowe}, \binits{D.M.}},
\bauthor{\bsnm{Sayle}, \binits{R.A.}},
\bauthor{\bsnm{Landrum}, \binits{G.A.}}:
\batitle{Development of a novel fingerprint for chemical reactions and its application to large-scale reaction classification and similarity}.
\bjtitle{Journal of chemical information and modeling}
\bvolume{55}(\bissue{1}),
\bfpage{39}--\blpage{53}
(\byear{2015})
\end{barticle}
\endbibitem

\bibitem[\protect\citeauthoryear{Karthikeyan et~al.}{2014}]{karthikeyan2014representation}
\begin{botherref}
\oauthor{\bsnm{Karthikeyan}, \binits{M.}},
\oauthor{\bsnm{Vyas}, \binits{R.}},
\oauthor{\bsnm{Karthikeyan}, \binits{M.}},
\oauthor{\bsnm{Vyas}, \binits{R.}}:
Representation, fingerprinting, and modelling of chemical reactions.
Practical Chemoinformatics,
317--374
(2014)
\end{botherref}
\endbibitem

\bibitem[\protect\citeauthoryear{Segler and Waller}{2017}]{segler2017neural}
\begin{barticle}
\bauthor{\bsnm{Segler}, \binits{M.H.}},
\bauthor{\bsnm{Waller}, \binits{M.P.}}:
\batitle{Neural-symbolic machine learning for retrosynthesis and reaction prediction}.
\bjtitle{Chemistry--A European Journal}
\bvolume{23}(\bissue{25}),
\bfpage{5966}--\blpage{5971}
(\byear{2017})
\end{barticle}
\endbibitem

\bibitem[\protect\citeauthoryear{McDermott et~al.}{2021}]{mcdermott2021graph}
\begin{barticle}
\bauthor{\bsnm{McDermott}, \binits{M.J.}},
\bauthor{\bsnm{Dwaraknath}, \binits{S.S.}},
\bauthor{\bsnm{Persson}, \binits{K.A.}}:
\batitle{A graph-based network for predicting chemical reaction pathways in solid-state materials synthesis}.
\bjtitle{Nature communications}
\bvolume{12}(\bissue{1}),
\bfpage{3097}
(\byear{2021})
\end{barticle}
\endbibitem

\bibitem[\protect\citeauthoryear{He et~al.}{2008}]{he2008graph}
\begin{barticle}
\bauthor{\bsnm{He}, \binits{K.}},
\bauthor{\bsnm{Ierapetritou}, \binits{M.G.}},
\bauthor{\bsnm{Androulakis}, \binits{I.P.}}:
\batitle{A graph-based approach to developing adaptive representations of complex reaction mechanisms}.
\bjtitle{Combustion and Flame}
\bvolume{155}(\bissue{4}),
\bfpage{585}--\blpage{604}
(\byear{2008})
\end{barticle}
\endbibitem

\bibitem[\protect\citeauthoryear{Jin et~al.}{2017}]{jin2017predicting}
\begin{botherref}
\oauthor{\bsnm{Jin}, \binits{W.}},
\oauthor{\bsnm{Coley}, \binits{C.}},
\oauthor{\bsnm{Barzilay}, \binits{R.}},
\oauthor{\bsnm{Jaakkola}, \binits{T.}}:
Predicting organic reaction outcomes with weisfeiler-lehman network.
Advances in neural information processing systems
\textbf{30}
(2017)
\end{botherref}
\endbibitem

\bibitem[\protect\citeauthoryear{Coley et~al.}{2019}]{coley2019graph}
\begin{barticle}
\bauthor{\bsnm{Coley}, \binits{C.W.}},
\bauthor{\bsnm{Jin}, \binits{W.}},
\bauthor{\bsnm{Rogers}, \binits{L.}},
\bauthor{\bsnm{Jamison}, \binits{T.F.}},
\bauthor{\bsnm{Jaakkola}, \binits{T.S.}},
\bauthor{\bsnm{Green}, \binits{W.H.}},
\bauthor{\bsnm{Barzilay}, \binits{R.}},
\bauthor{\bsnm{Jensen}, \binits{K.F.}}:
\batitle{A graph-convolutional neural network model for the prediction of chemical reactivity}.
\bjtitle{Chemical science}
\bvolume{10}(\bissue{2}),
\bfpage{370}--\blpage{377}
(\byear{2019})
\end{barticle}
\endbibitem

\bibitem[\protect\citeauthoryear{Irwin et~al.}{2022}]{irwin2022chemformer}
\begin{barticle}
\bauthor{\bsnm{Irwin}, \binits{R.}},
\bauthor{\bsnm{Dimitriadis}, \binits{S.}},
\bauthor{\bsnm{He}, \binits{J.}},
\bauthor{\bsnm{Bjerrum}, \binits{E.J.}}:
\batitle{Chemformer: a pre-trained transformer for computational chemistry}.
\bjtitle{Machine Learning: Science and Technology}
\bvolume{3}(\bissue{1}),
\bfpage{015022}
(\byear{2022})
\end{barticle}
\endbibitem

\bibitem[\protect\citeauthoryear{Jiang et~al.}{2021}]{jiang2021smiles}
\begin{barticle}
\bauthor{\bsnm{Jiang}, \binits{S.}},
\bauthor{\bsnm{Zhang}, \binits{Z.}},
\bauthor{\bsnm{Zhao}, \binits{H.}},
\bauthor{\bsnm{Li}, \binits{J.}},
\bauthor{\bsnm{Yang}, \binits{Y.}},
\bauthor{\bsnm{Lu}, \binits{B.-L.}},
\bauthor{\bsnm{Xia}, \binits{N.}}:
\batitle{When smiles smiles, practicality judgment and yield prediction of chemical reaction via deep chemical language processing}.
\bjtitle{IEEE Access}
\bvolume{9},
\bfpage{85071}--\blpage{85083}
(\byear{2021})
\end{barticle}
\endbibitem

\bibitem[\protect\citeauthoryear{Schwaller et~al.}{2021}]{schwaller2021prediction}
\begin{barticle}
\bauthor{\bsnm{Schwaller}, \binits{P.}},
\bauthor{\bsnm{Vaucher}, \binits{A.C.}},
\bauthor{\bsnm{Laino}, \binits{T.}},
\bauthor{\bsnm{Reymond}, \binits{J.-L.}}:
\batitle{Prediction of chemical reaction yields using deep learning}.
\bjtitle{Machine learning: science and technology}
\bvolume{2}(\bissue{1}),
\bfpage{015016}
(\byear{2021})
\end{barticle}
\endbibitem

\bibitem[\protect\citeauthoryear{Schwaller et~al.}{2018}]{schwaller2018found}
\begin{barticle}
\bauthor{\bsnm{Schwaller}, \binits{P.}},
\bauthor{\bsnm{Gaudin}, \binits{T.}},
\bauthor{\bsnm{Lanyi}, \binits{D.}},
\bauthor{\bsnm{Bekas}, \binits{C.}},
\bauthor{\bsnm{Laino}, \binits{T.}}:
\batitle{“found in translation”: predicting outcomes of complex organic chemistry reactions using neural sequence-to-sequence models}.
\bjtitle{Chemical science}
\bvolume{9}(\bissue{28}),
\bfpage{6091}--\blpage{6098}
(\byear{2018})
\end{barticle}
\endbibitem

\bibitem[\protect\citeauthoryear{Karpov et~al.}{2019}]{karpov2019transformer}
\begin{bchapter}
\bauthor{\bsnm{Karpov}, \binits{P.}},
\bauthor{\bsnm{Godin}, \binits{G.}},
\bauthor{\bsnm{Tetko}, \binits{I.V.}}:
\bctitle{A transformer model for retrosynthesis}.
In: \bbtitle{International Conference on Artificial Neural Networks},
pp. \bfpage{817}--\blpage{830}
(\byear{2019}).
\bcomment{Springer}
\end{bchapter}
\endbibitem

\bibitem[\protect\citeauthoryear{Weininger}{1988}]{weininger1988smiles}
\begin{barticle}
\bauthor{\bsnm{Weininger}, \binits{D.}}:
\batitle{Smiles, a chemical language and information system. 1. introduction to methodology and encoding rules}.
\bjtitle{Journal of chemical information and computer sciences}
\bvolume{28}(\bissue{1}),
\bfpage{31}--\blpage{36}
(\byear{1988})
\end{barticle}
\endbibitem

\bibitem[\protect\citeauthoryear{Mislow}{2012}]{mislow2012introduction}
\begin{bbook}
\bauthor{\bsnm{Mislow}, \binits{K.}}:
\bbtitle{Introduction to Stereochemistry}.
\bpublisher{Courier Corporation}, \blocation{???}
(\byear{2012})
\end{bbook}
\endbibitem

\bibitem[\protect\citeauthoryear{N{\'o}gr{\'a}di et~al.}{2016}]{nogradi2016stereochemistry}
\begin{bbook}
\bauthor{\bsnm{N{\'o}gr{\'a}di}, \binits{M.}},
\bauthor{\bsnm{Poppe}, \binits{L.}},
\bauthor{\bsnm{Nagy}, \binits{J.}},
\bauthor{\bsnm{Horny{\'a}nszky}, \binits{G.}},
\bauthor{\bsnm{Boros}, \binits{Z.}}:
\bbtitle{Stereochemistry and Stereoselective Synthesis: An Introduction}.
\bpublisher{John Wiley \& Sons}, \blocation{???}
(\byear{2016})
\end{bbook}
\endbibitem

\bibitem[\protect\citeauthoryear{Andersen et~al.}{2017}]{andersen2017chemical}
\begin{bchapter}
\bauthor{\bsnm{Andersen}, \binits{J.L.}},
\bauthor{\bsnm{Flamm}, \binits{C.}},
\bauthor{\bsnm{Merkle}, \binits{D.}},
\bauthor{\bsnm{Stadler}, \binits{P.F.}}:
\bctitle{Chemical graph transformation with stereo-information}.
In: \bbtitle{Graph Transformation: 10th International Conference, ICGT 2017, Held as Part of STAF 2017, Marburg, Germany, July 18-19, 2017, Proceedings 10},
pp. \bfpage{54}--\blpage{69}
(\byear{2017}).
\bcomment{Springer}
\end{bchapter}
\endbibitem

\bibitem[\protect\citeauthoryear{Cao et~al.}{2024}]{cao2024presto}
\begin{bchapter}
\bauthor{\bsnm{Cao}, \binits{H.}},
\bauthor{\bsnm{Shao}, \binits{Y.}},
\bauthor{\bsnm{Liu}, \binits{Z.}},
\bauthor{\bsnm{Liu}, \binits{Z.}},
\bauthor{\bsnm{Tang}, \binits{X.}},
\bauthor{\bsnm{Yao}, \binits{Y.}},
\bauthor{\bsnm{Li}, \binits{Y.}}:
\bctitle{{PRESTO}: Progressive pretraining enhances synthetic chemistry outcomes}.
In: \beditor{\bsnm{Al-Onaizan}, \binits{Y.}},
\beditor{\bsnm{Bansal}, \binits{M.}},
\beditor{\bsnm{Chen}, \binits{Y.-N.}} (eds.)
\bbtitle{Findings of the Association for Computational Linguistics: EMNLP 2024},
pp. \bfpage{10197}--\blpage{10224}.
\bpublisher{Association for Computational Linguistics},
\blocation{Miami, Florida, USA}
(\byear{2024}).
\burl{https://aclanthology.org/2024.findings-emnlp.597}
\end{bchapter}
\endbibitem

\bibitem[\protect\citeauthoryear{Mentzer et~al.}{2024}]{mentzer2024finite}
\begin{bchapter}
\bauthor{\bsnm{Mentzer}, \binits{F.}},
\bauthor{\bsnm{Minnen}, \binits{D.}},
\bauthor{\bsnm{Agustsson}, \binits{E.}},
\bauthor{\bsnm{Tschannen}, \binits{M.}}:
\bctitle{Finite scalar quantization: {VQ}-{VAE} made simple}.
In: \bbtitle{The Twelfth International Conference on Learning Representations}
(\byear{2024}).
\burl{https://openreview.net/forum?id=8ishA3LxN8}
\end{bchapter}
\endbibitem

\bibitem[\protect\citeauthoryear{Sterling and Irwin}{2015}]{sterling2015zinc}
\begin{barticle}
\bauthor{\bsnm{Sterling}, \binits{T.}},
\bauthor{\bsnm{Irwin}, \binits{J.J.}}:
\batitle{Zinc 15--ligand discovery for everyone}.
\bjtitle{Journal of chemical information and modeling}
\bvolume{55}(\bissue{11}),
\bfpage{2324}--\blpage{2337}
(\byear{2015})
\end{barticle}
\endbibitem

\bibitem[\protect\citeauthoryear{Gaulton et~al.}{2012}]{gaulton2012chembl}
\begin{barticle}
\bauthor{\bsnm{Gaulton}, \binits{A.}},
\bauthor{\bsnm{Bellis}, \binits{L.J.}},
\bauthor{\bsnm{Bento}, \binits{A.P.}},
\bauthor{\bsnm{Chambers}, \binits{J.}},
\bauthor{\bsnm{Davies}, \binits{M.}},
\bauthor{\bsnm{Hersey}, \binits{A.}},
\bauthor{\bsnm{Light}, \binits{Y.}},
\bauthor{\bsnm{McGlinchey}, \binits{S.}},
\bauthor{\bsnm{Michalovich}, \binits{D.}},
\bauthor{\bsnm{Al-Lazikani}, \binits{B.}}, \betal:
\batitle{Chembl: a large-scale bioactivity database for drug discovery}.
\bjtitle{Nucleic acids research}
\bvolume{40}(\bissue{D1}),
\bfpage{1100}--\blpage{1107}
(\byear{2012})
\end{barticle}
\endbibitem

\bibitem[\protect\citeauthoryear{Liu et~al.}{2017}]{liu2017retrosynthetic}
\begin{barticle}
\bauthor{\bsnm{Liu}, \binits{B.}},
\bauthor{\bsnm{Ramsundar}, \binits{B.}},
\bauthor{\bsnm{Kawthekar}, \binits{P.}},
\bauthor{\bsnm{Shi}, \binits{J.}},
\bauthor{\bsnm{Gomes}, \binits{J.}},
\bauthor{\bsnm{Luu~Nguyen}, \binits{Q.}},
\bauthor{\bsnm{Ho}, \binits{S.}},
\bauthor{\bsnm{Sloane}, \binits{J.}},
\bauthor{\bsnm{Wender}, \binits{P.}},
\bauthor{\bsnm{Pande}, \binits{V.}}:
\batitle{Retrosynthetic reaction prediction using neural sequence-to-sequence models}.
\bjtitle{ACS central science}
\bvolume{3}(\bissue{10}),
\bfpage{1103}--\blpage{1113}
(\byear{2017})
\end{barticle}
\endbibitem

\bibitem[\protect\citeauthoryear{Wang et~al.}{2023}]{wang2023generic}
\begin{barticle}
\bauthor{\bsnm{Wang}, \binits{X.}},
\bauthor{\bsnm{Hsieh}, \binits{C.-Y.}},
\bauthor{\bsnm{Yin}, \binits{X.}},
\bauthor{\bsnm{Wang}, \binits{J.}},
\bauthor{\bsnm{Li}, \binits{Y.}},
\bauthor{\bsnm{Deng}, \binits{Y.}},
\bauthor{\bsnm{Jiang}, \binits{D.}},
\bauthor{\bsnm{Wu}, \binits{Z.}},
\bauthor{\bsnm{Du}, \binits{H.}},
\bauthor{\bsnm{Chen}, \binits{H.}}, \betal:
\batitle{Generic interpretable reaction condition predictions with open reaction condition datasets and unsupervised learning of reaction center}.
\bjtitle{Research}
\bvolume{6},
\bfpage{0231}
(\byear{2023})
\end{barticle}
\endbibitem

\bibitem[\protect\citeauthoryear{Levenshtein et~al.}{1966}]{levenshtein1966binary}
\begin{bchapter}
\bauthor{\bsnm{Levenshtein}, \binits{V.I.}}, \betal:
\bctitle{Binary codes capable of correcting deletions, insertions, and reversals}.
In: \bbtitle{Soviet Physics Doklady},
vol. \bseriesno{10},
pp. \bfpage{707}--\blpage{710}
(\byear{1966}).
\bcomment{Soviet Union}
\end{bchapter}
\endbibitem

\bibitem[\protect\citeauthoryear{Riniker and Landrum}{2015}]{riniker2015better}
\begin{barticle}
\bauthor{\bsnm{Riniker}, \binits{S.}},
\bauthor{\bsnm{Landrum}, \binits{G.A.}}:
\batitle{Better informed distance geometry: using what we know to improve conformation generation}.
\bjtitle{Journal of chemical information and modeling}
\bvolume{55}(\bissue{12}),
\bfpage{2562}--\blpage{2574}
(\byear{2015})
\end{barticle}
\endbibitem

\bibitem[\protect\citeauthoryear{Halgren}{1996}]{halgren1996merck}
\begin{barticle}
\bauthor{\bsnm{Halgren}, \binits{T.A.}}:
\batitle{Merck molecular force field. i. basis, form, scope, parameterization, and performance of mmff94}.
\bjtitle{Journal of computational chemistry}
\bvolume{17}(\bissue{5-6}),
\bfpage{490}--\blpage{519}
(\byear{1996})
\end{barticle}
\endbibitem

\bibitem[\protect\citeauthoryear{Raffel et~al.}{2020}]{raffel2020exploring}
\begin{barticle}
\bauthor{\bsnm{Raffel}, \binits{C.}},
\bauthor{\bsnm{Shazeer}, \binits{N.}},
\bauthor{\bsnm{Roberts}, \binits{A.}},
\bauthor{\bsnm{Lee}, \binits{K.}},
\bauthor{\bsnm{Narang}, \binits{S.}},
\bauthor{\bsnm{Matena}, \binits{M.}},
\bauthor{\bsnm{Zhou}, \binits{Y.}},
\bauthor{\bsnm{Li}, \binits{W.}},
\bauthor{\bsnm{Liu}, \binits{P.J.}}:
\batitle{Exploring the limits of transfer learning with a unified text-to-text transformer}.
\bjtitle{Journal of machine learning research}
\bvolume{21}(\bissue{140}),
\bfpage{1}--\blpage{67}
(\byear{2020})
\end{barticle}
\endbibitem

\end{thebibliography}

\end{document}